\documentclass[aps,twocolumn,secnumarabic,balancelastpage,amsmath,amssymb,longbibliography]{revtex4-2}

\usepackage{amsmath}   
\usepackage{graphicx}  
\usepackage{textcomp}
\usepackage{graphicx}
\usepackage[dvipsnames]{xcolor}
\usepackage[colorlinks,linkcolor=black,citecolor=black,urlcolor=black,filecolor=black]{hyperref}
\usepackage{amsmath, amssymb, amsthm}
\usepackage{wasysym}
\usepackage{lipsum}
\usepackage{outlines}
\usepackage{enumerate}
\usepackage{setspace}
\usepackage{tabularx} 
\usepackage{array}
\newcolumntype{P}[1]{>{\centering\arraybackslash}p{#1}}
\usepackage{xcolor}
\usepackage[export]{adjustbox}
\usepackage{tikz}
\usepackage{verbatim}
\usetikzlibrary{decorations.pathreplacing}
\usetikzlibrary{arrows,arrows.meta,calc,shapes.geometric,shapes.misc}
\tikzset{
	>=stealth',
	help lines/.style={dashed, thick},
	important line/.style={thick},
	connection/.style={thick, dotted},
}
\DeclareMathAlphabet{\mymathbb}{U}{BOONDOX-ds}{m}{n}
\usepackage{dsfont}
\usepackage{braket}
\hypersetup{
    pdfstartview={FitH}, 
    colorlinks=true, 
    linkcolor=blue, 
    citecolor=blue, 
    filecolor=magenta,
    urlcolor=blue
}
\usepackage{algorithm}
\usepackage{algpseudocode}

\usepackage{wrapfig}

\usepackage{algorithm}
\usepackage{algpseudocode}
\usepackage{verbatim}

\usepackage[normalem]{ulem}

\usepackage{xcolor}

\begin{document}

\title{Leveraging Qubit Loss Detection in Fault Tolerant Quantum Algorithms}

\author{
Gefen~Baranes$^{1,2,*, \dagger}$, Madelyn~Cain$^{2,*}$, J. Pablo Bonilla~Ataides$^{2,*}$, Dolev~Bluvstein$^{2}$, Josiah~Sinclair$^{1}$, Vladan~Vuleti\ifmmode \acute{c}\else \'{c}\fi{}$^{1}$, Hengyun~Zhou$^{2,3}$, and Mikhail~D.~Lukin$^{2,\ddagger}$}
\affiliation{
$^1$Department~of~Physics~and~Research~Laboratory~of~Electronics,~Massachusetts~Institute~of~Technology,~Cambridge, MA, USA \\ 
$^2$Department~of~Physics,~Harvard~University,~Cambridge,~MA~02138,~USA\\ 
$^3$QuEra Computing Inc.,~Boston,~MA~02135,~USA\\
$^*$These~authors~contributed~equally;~$^\dagger$gbaranes@mit.edu;~$^\ddagger$lukin@physics.harvard.edu
}

\date{\today}

\begin{abstract}
Qubit loss errors constitute a dominant source of noise in many quantum hardware systems, particularly in neutral atom quantum computers.
We develop a theoretical framework to effectively detect and correct loss errors in logical algorithms and leverage such loss information in decoding. 
Considering general quantum error correction codes and logical circuits, we introduce a delayed-erasure decoder for experimentally-motivated 
error models which leverages information from  delayed loss detection to accurately correct loss errors, even when the precise moment of the error is unknown.
Using this decoder, we identify strategies for detecting and correcting loss errors based on the logical circuit structure.
For deep circuits prior to logical measurement, we explore methods to integrate loss detection into syndrome extraction with minimal overhead, identifying optimal strategies depending on the qubit loss fraction in the noise and hardware capabilities.
In contrast, we find that many key algorithmic subroutines involve frequent gate teleportation, shortening the circuit depth before logical measurement and naturally replacing qubits with no additional experimental overhead.
We simulate this setting using a toy model algorithm for small-angle synthesis, 
and find a significant performance improvement as the loss fraction increases.
These results provide a path forward for advancing large-scale fault tolerant quantum computation in systems with loss error detection.

\end{abstract}

\maketitle

\section{Introduction}

Quantum error correction (QEC) is believed to be essential for realizing large-scale quantum computation, as it enables suppression of errors~\cite{shor1995scheme, calderbank1996good, shor1996fault}.
However, its practical implementation remains challenging due to its substantial resource overhead.
Recent experimental advancements have demonstrated remarkable progress in implementing quantum operations across multiple logical qubits~\cite{bluvstein2024logical, google2023suppressing, AWS2024hardware, Quantinuum2024demonstration, reichardt2024logical}, and operating below error thresholds~\cite{google2024quantum, bluvstein2025architectural}. 
These advances make it clear that practical QEC performance can be substantially improved by tailoring the error correction strategy to the particular experimental error model~\cite{ma2023high, google2024quantum, scholl2023erasure}, choice of logic gates~\cite{bluvstein2024logical}, and the structure of the algorithm itself~\cite{cain2024correlated, zhou2024algorithmic}.

In particular, qubit loss and leakage from computational subspace are dominant noise sources in many hardware systems, and can have a dramatic effect on the practical performance of quantum processors \cite{evered2023high, miao2023overcoming}.
In the absence of loss correction, all qubits within a QEC code can eventually disappear, destroying the encoded quantum information. Conversely, directly detecting loss provides valuable information about the error location, in contrast with Pauli errors, which are indirectly inferred from syndrome information. 
Recent work has shown that the rich information provided by so-called erasure detection can in fact improve QEC performance substantially~\cite{sahay2023high, wu2022erasure, kang2023quantum, kubica2023erasure, ma2023high, scholl2023erasure, chang2024surface, gu2023fault, yu2024tracking, perrin2024quantum, chow2024SSR, cong2022hardware, Riverlane2024local}. 
Similarly, noise bias present in many platforms~\cite{evered2023high, bluvstein2024logical,google2024quantum, bluvstein2025architectural} can also be leveraged to improve performance~\cite{nelson2024assessment, bonilla2021xzzx, darmawan2021practical, claes2023tailored}. 
These studies motivate developing strategies that incorporate loss detection and harness bias to improve practical QEC performance.

In this Article, we explore the role of loss errors in error-corrected circuits across a range of quantum hardware platforms, including alkali and alkaline-earth-like neutral atoms, as well as superconducting qubits or trapped ions.
We focus on the realistic scenario in which detection and correction of loss are delayed for several gate operations, and develop techniques that leverage this delayed information with minimal experimental overhead.
In particular, we develop a delayed-erasure decoder to accurately interpret logical measurement results from the measured syndromes and loss detections, despite uncertainty in the exact moment of the loss error.

Using this delayed-erasure decoder, we investigate the impact of loss errors in logical circuits.
We demonstrate that the algorithmic structure significantly influences the optimal strategy to detect and correct loss, as summarized in Fig.~\ref{fig:illustration_logical_alg}(a). 
Concretely, for high-depth circuits prior to logical measurement, we develop hardware-efficient methods to detect and replace lost qubits during syndrome extraction (SE) with minimal additional overhead.
We perform circuit-level simulations to compare modified circuit-based~\cite{shor1996fault} and measurement-based~\cite{raussendorf2003measurement}
approaches which incorporate loss detection upon measurement, alongside other state-of-the-art approaches such as erasure conversion techniques~\cite{ma2023high, wu2022erasure}.
We observe that QEC performance can be substantially improved by optimizing circuit design to leverage loss errors in cases where it constitutes a substantial fraction of the error budget.
By analyzing biased errors we find that loss has a stronger impact on performance than bias, even with bias-preserving gates.
We then study how loss-detecting SE rounds can be optimally interleaved between transversal gates in multi-qubit deep logical Clifford circuits.

We next find that many key algorithmic subroutines have short circuit depth before logical measurement, including magic state distillation~\cite{bravyi2005universal}, quantum arithmetic~\cite{gidney2018adder}, and small-angle synthesis~\cite{kitaev1997quantum, kitaev2002classical}. In such circuits, we find that because gate teleportation naturally detects and replaces lost qubits, no loss-detecting SE is needed.
We discover through numerical simulations of a teleportation-based algorithm that as the frequency of gate teleportation increases, the performance
approaches that of frequent erasure conversion.
These results provide a comparative analysis of experimental solutions for leveraging loss, and highlight that algorithmic structure plays a central role in how these factors impact performance.

Before proceeding, we note that loss-to-erasure conversion has been explored at the memory level, using mid-circuit measurement in alkaline-earth-like atomic systems \cite{wu2022erasure, ma2023high, sahay2023high, scholl2023erasure} and superconducting qubits \cite{kubica2023erasure, gu2023fault, chang2024surface, miao2023overcoming, levine2024demonstrating, de2025mid, chou2023demonstrating}, as well as using leakage-reduction units with extra qubits and gates \cite{cong2022hardware, fowler2013coping, aliferis2005fault, suchara2015leakage}.
More recently, loss-resolving readout was demonstrated experimentally in alkali atoms systems and dual-rail superconducting qubits and its implications for logical qubit performance were explored in 
Refs. \cite{yu2024tracking, chow2024SSR,chou2024superconducting,  perrin2024quantum,reichardt2024logical, bluvstein2025architectural}.
Here we present an end-to-end analysis of loss handling techniques, comparing all these approaches across a range of quantum computing platforms.
Our study extends beyond single logical qubits to examine the role of loss at every level of computation, from decoding strategies and QEC architectures to full logical algorithms.
We systematically compare different loss detection and correction methods, including their frequencies and associated space-time overheads.
By focusing on logical algorithms, we provide new insights into how circuit structure influences the algorithmic performance and develop new, optimized strategies for circumventing loss errors.

Our paper is organized as follows. 
Section~\ref{sec:detecting_decoding_loss} discusses the effect of qubit loss and its detection using state-selective readout, and the delayed-erasure decoder developed in this work.
In Section~\ref{sec:SE_techniques}, we tailor and compare different SE techniques designed to handle loss with minimal overhead, which are necessary in circuits with high depth before logical measurements.
Section~\ref{sec:loss_random_algs} explores the optimal frequency of interleaving loss-detecting SE with transversal gates in deep logical Clifford circuits.
In Section~\ref{sec:logical_algs_with_teleportation}, we highlight the prevalent use of teleportation in many algorithmic subroutines and evaluate its effect with and without the delayed-erasure decoder on a toy model for the small-angle synthesis algorithm.
Finally, we present our conclusions in Section~\ref{sec:conclusion}.

\section{Detecting and decoding delayed erasures}
\label{sec:detecting_decoding_loss}
\begin{figure}[t]
  \centering
  \includegraphics[width=\columnwidth]{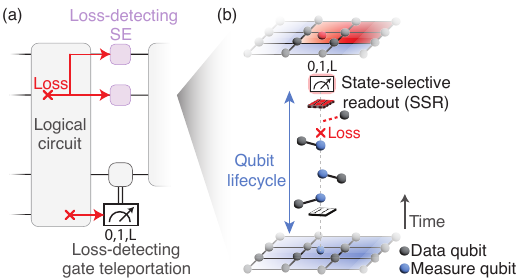}
  \caption{Loss errors in logical circuits. (a)~Depiction of a logical algorithm with loss-detecting SE and gate teleportation. Physical qubit losses (red crosses) can generate correlated errors within and between logical qubits. (b)~Space-time diagram of a logical circuit, focusing on a measure qubit lifecycle during syndrome extraction. Physical qubits progress through time, undergoing initialization, gate operations, idling, and measurement. A loss event causes future gates to be canceled, generating correlated errors between the qubits in the gate and flipping the corresponding stabilizers. 
  }
  \label{fig:illustration_logical_alg}
\end{figure}

We consider general logical algorithms subject to loss errors, in which loss is periodically detected using hardware-specific mechanisms. 
In alkali neutral atom systems or dual rail superconducting qubits, this can be done via \textbf{state-selective readout (SSR)}, a projective measurement that distinguishes between $\ket{0}$, $\ket{1}$, and loss.
As it occurs as part of the qubit measurement, it can be realized through various methods with minimal experimental overhead~\cite{deist2022mid, hu2024site, scholl2023erasure, lis2023midcircuit, graham2023midcircuit, huie2023SSR, chow2024SSR, reichardt2024logical, chou2024superconducting, bluvstein2025architectural}.
In contrast, \textbf{mid-circuit erasure conversion}, demonstrated in alkaline-earth-like atoms and superconducting qubits~\cite{wu2022erasure, google2023sparse, koottandavida2024erasure}, identifies loss or leakage events during circuit execution by monitoring auxiliary degrees of freedom without performing a full projective measurement.
These detection mechanisms, combined with methods like teleportation-based protocols, feed into our syndrome extraction and decoding framework (see Fig.~\ref{fig:illustration_logical_alg}(a)).

We explicitly differentiate between two closely related error types. 
We define a \textbf{loss} error as one in which a qubit leaves the computational subspace and becomes unavailable until detection and replacement, encompassing both loss and leakage (depending on hardware platform). 
An \textbf{erasure} error, by contrast, is a loss that is detected at the gate layer where it occurs and is immediately reinitialized; both the time and qubit index are known to the decoder.

Loss events lead to correlated errors: in neutral-atom systems, a lost qubit cancels subsequent gates, while its intended interaction partner may experience an effective single-qubit error~\cite{evered2023high}; 
in superconducting platforms, 
a qubit interacting with a lost qubit suffers from depolarizing noise~\cite{chang2024surface}.
In both cases, correlated errors can propagate within or across logical qubits during entangling operations (see Fig.~\ref{fig:illustration_logical_alg}(a,b)), and must be handled carefully by the decoder.
Here we adopt the gate-cancellation model for loss.

\begin{figure}[t]
  \centering
  \includegraphics[width=\columnwidth]{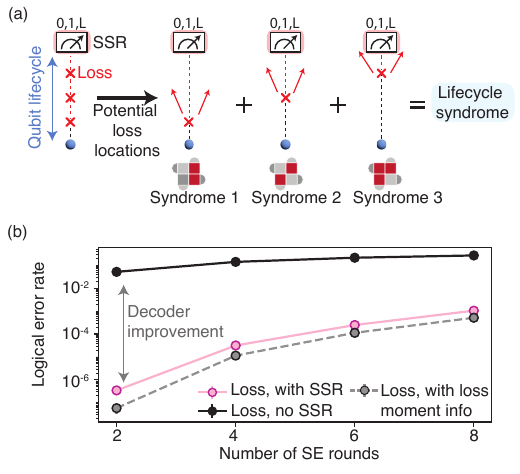}
  \caption{Delayed-erasure decoder. (a) Illustration of a qubit lifecycle and its usage in the delayed-erasure decoder. 
  From initialization to measurement, each physical qubit can be lost at multiple possible time points, each occurring with a potentially different probability and corresponding syndrome. 
  Upon detection of a qubit loss, the decoder accounts for these possibilities in order to improve the accuracy of the assigned correction.
  (b) Logical error rate for a logical memory as a function of the number of conventional SE rounds before logical measurement, here with distance $d=5$ and loss errors with probability $p_{\text{loss}}=1\%$ per entangling gate. 
  The delayed-erasure decoder (pink) substantially outperforms a decoder which does not account for loss information (black).
  Even without loss moment information, the delayed-erasure decoder achieves comparable performance to a decoder with perfect loss time-location (gray, dashed).
  }
  \label{fig:Delayed_erasure_decoder}
\end{figure}

While in principle these errors can be corrected by detecting and replacing the lost qubits and treating them as erasure errors,
unlike the ideal erasure channels, where the moment in time of the loss error is precisely known, in a natural QEC cycle, the loss event is detected when the SSR or other loss detection mechanism is performed and can correspond to a number of different potential loss locations (Fig.~\ref{fig:Delayed_erasure_decoder}(a)),
each resulting in a potentially different set of correlated errors.
These loss detection events can be viewed as \textit{delayed} erasure detections, and the information revealed by the detection mechanism can be leveraged by the appropriate decoder to improve logical performance.

As a result, loss events require two adjustments to the decoder: first, when a qubit is lost, its associated stabilizer checks are no longer valid. Therefore, they need to be replaced with a so-called ``supercheck", a product of multiple stabilizer checks into a single check which is independent of the lost qubit~\cite{stace2009thresholds, barrett2010fault}.
For example, consider a lost qubit $q_4$ that participates in neighboring stabilizers $S_1 = Z_1 Z_2 Z_3 Z_4$ and $S_2 = Z_4 Z_5 Z_6 Z_7$. The resulting supercheck operator is $S_1 S_2 = Z_1 Z_2 Z_3 Z_5 Z_6 Z_7$, which is independent of the lost qubit $q_4$.
Second, loss causes correlated errors, necessitating updates to the circuit error model to account for the likelihood of error propagation from the possible loss locations (see Appendix Section~\ref{app:syndrome_of_loss_errors}).

We develop a delayed-erasure decoder which includes these two adjustments to effectively utilize imperfect information about the location of losses in time obtained from SSR. 
As a key quantity in predicting the performance of different loss detection and correction methods, we consider the qubit lifecycle, defined as the number of circuit locations where a given qubit can potentially be lost, starting at initialization and ending at measurement.
Formalizing our approach, our goal is to automatically construct a decoding hypergraph based on the observed loss, which captures how errors (hyperedges) trigger checks (vertices that compare consecutive stabilizer checks in time) 
and is the input to the decoder.
Ideally, we would solve the most-likely-error (MLE) decoding problem, which identifies the most likely configuration of Pauli and loss errors consistent with both the observed error syndromes and loss events (see Appendix Section~B.\ref{appendix:mle_decoder}).
However, a full MLE solution would require considering all combinations of loss locations, since loss-induced errors can produce non-Pauli, correlated error patterns, making them non-additive, as observed numerically.

Instead, we approximate the MLE solution by handling each loss event independently 
(see Appendix Section~B\ref{appendix:approx-mle} for full details).
For each loss event, we trace back the qubit’s lifecycle, accounting for all potential loss events and their associated probabilities. 
Each potential loss event corresponds to a loss circuit, in which certain gates are canceled due to the loss. Using Stim~\cite{gidney2021stim}, a Clifford circuit simulator, we construct the hyperedges and their probabilities for each loss circuit. 
Next, we integrate all hyperedges within a qubit’s lifecycle, re-weighting them based on the probabilities of their corresponding potential loss locations.
In cases of multiple losses, each loss event is calculated independently, and the results are combined to construct the final decoding hypergraph used by the decoder, calculated by:
\begin{equation}
\sum_i {D}_i + {D}_{\mathrm{Pauli}} + \omega \cdot {D}_{\mathrm{first\ comb}},
\end{equation}
where $D_i$ is the decoding hypergraph for a lossy lifecycle $i$, $D_{\mathrm{Pauli}}$ is the decoding hypergraph for Pauli errors generated from the lossless circuit, and $D_{\mathrm{first\ comb}}$ is the decoding hypergraph built from the earliest potential loss in each lossy lifecycle $i$.
The weighting parameter $\omega$ tunes the contribution of $D_{\mathrm{first\ comb}}$, and is set to $\omega = 0$ in the main-text simulations (see Appendix~\ref{appendix:delayed_erasure_decoder}.2 for details).
Empirically, this heuristic achieves performance comparable to methods that consider combinations of loss events, while significantly reducing computational overhead.

The updated decoding hypergraph can then be processed using a standard decoder appropriate for the QEC code in use.
For example, one may use a MLE decoder, minimum weight perfect matching (MWPM) for topological codes~\cite{google2023sparse, wu2023fusion, google2024quantum}, belief propagation with ordered statistics decoding (BP+OSD) for sparse codes~\cite{panteleev2021degenerate}, or machine learning-based decoders~\cite{google2024learning, google2024quantum}. 
Unlike previous approaches to decoding loss errors~\cite{gu2023fault, google2024quantum, yu2024tracking}, our decoder automatically adjusts the error model based on any general circuit and loss information, eliminating the need for hand-tuned models that may not easily generalize to complex logical algorithms or different SE methods.
Here we primarily use the MLE decoder from Ref.~\cite{cain2024correlated}, as it can decode generic stabilizer codes and logical algorithms. We also present results for the MWPM decoder in Appendix~Section~\ref{app:interplay_loss_erasure_biased_errors}.

To benchmark the decoder, we use a surface-code logical memory with repeated SE rounds in which loss is detected only by SSR: losses on measure qubits are detected each round, whereas losses on data qubits are detected only at the final projective readout. 
Unless otherwise noted, this paper considers loss and Pauli errors occurring during entangling operations, as this represents a dominant source of error in many quantum computing architectures~\cite{bluvstein2024logical, google2024quantum, evered2023high}. 
Details of the error model, along with results using alternative error models, can be found in Appendix~\ref{supp:error_models_loss} and Table~\ref{table:syndrome_comparison}.
As shown in Fig.~\ref{fig:Delayed_erasure_decoder}(b), a delayed-erasure decoder that uses only SSR information lowers the logical error rate by orders of magnitude relative to an MLE decoder that ignores loss information and remains comparable to a decoder with perfect loss time-location, even for deeper circuits with longer lifecycles.

\section{Techniques for addressing qubit loss in deep circuits}
\label{sec:SE_techniques}

In deep logical circuits prior to logical measurement, qubit lifecycles are extended, introducing many correlated errors and degrading performance. 
While the delayed-erasure decoder mitigates these effects, its effectiveness deteriorates when lifecycles become too long due to error accumulation.
To maintain high performance, loss detection and qubit replacement must be integrated into the QEC process. 
In this section, we explore practical syndrome extraction (SE) methods that enable such replenishment with minimal overhead.
We primarily focus on: (i) conventional SE, which resembles traditional circuit-based quantum computing (CBQC) with various modifications for delayed-erasure conversion utilizing SSR or erasure conversion capabilities, and (ii) teleportation-based SE, which resembles measurement-based quantum computing (MBQC)~\cite{raussendorf2001one, raussendorf2002one}.
Additionally, a description of a modified Steane SE approach, which bridges conventional and teleportation-based SE but is not simulated here, is provided in Appendix Section~\ref{app:SE_methods_info}.\ref{app:Steane_SE}.
These methods differ in their qubit overhead, gate operations, loss detection capabilities, and experimental requirements, as analyzed below.

We find that the underlying circuits and resulting performance of the methods are similar, but not identical: all approaches successfully remove qubit loss and have trade-offs in logical error rates and qubit overheads depending on the specific ratios between loss and Pauli errors.
We begin by detailing each SE approach before analyzing numerical results in Section~\ref{sec:comparing_different_SE}.
In Section~\ref{subsec:linking_performance_to_key_metrics}, we relate these performance differences to key metrics such as lifecycle length and number of entangling gates to predict performance based on the noise model for each SE method.

\subsection{\textit{Modified conventional SE}}
\label{subsection:SWAP_SE}
The conventional SE method involves repeated stabilizer checks using physical measure qubits. 
Data qubits are not directly measured or replaced when lost, and thus over time, logical performance can degrade in the presence of loss.
We consider augmenting conventional SE by utilizing SSR and physical SWAP gates to detect losses on all qubits (SWAP SE), as proposed in Refs.~\cite{ghosh2013understanding, suchara2015leakage} and further explored in Refs.~\cite{chow2024SSR, perrin2024quantum}. 
At the end of each SE round, a SWAP gate and physical SWAP movement are performed between data and measure qubits. 
Conveniently, this approach does not require any additional entangling gates by applying gate cancellation identities (see Fig.~\ref{fig:syndrome_extraction_methods_memory}(a)).
If the data qubit is not lost, the SWAP operations cancel each other, such that the resulting measure qubit stores the stabilizer outcome.
Conversely, if the data qubit was lost, the loss is directly identified through the SSR measurement and automatically replaced.
Thus, using this method, we correct losses of all types of qubits, each cycles through both roles of data and measure qubits,
ensuring a uniform lifecycle length of $\sim 8$ for all qubits in the bulk even in deep circuits (see Appendix~\ref{app:SWAP_SE}).
However, a data qubit loss detection comes at the cost of losing a stabilizer outcome information, potentially impacting performance, as observed numerically in the next section.

Each SWAP operation incurs a cost due to qubit movement, making it crucial to evaluate different SWAP periods to balance performance and experimental complexity.
The SWAP period defines how often loss-detecting SE rounds are interspersed with conventional SE rounds, occurring at a fixed interval.
We identify a trade-off between the lifecycle lengths of data and measurement qubits for different SWAP periods, demonstrating that a period of 2 can be competitive with a period of 1 (see Figures.~\ref{supp_fig:SWAP_freq12_lifecycles},~\ref{app_fig:optimal_SWAP_period}). 
A period of 2 reduces experimental complexity while maintaining a similar average lifecycle length.

\subsection{\textit{Teleportation-based SE}}

Teleportation-based SE uses repeated logical teleportation to fresh code blocks, inherently detecting loss via SSR at each step~\cite{raussendorf2001one,raussendorf2002one,yu2024tracking,sahay2023high} (see Fig.~\ref{fig:syndrome_extraction_methods_memory}(b)). 
In our implementation, logical qubits are prepared in one initialization round and then entangled in the XZZX pattern, which is equivalent to the XZZX cluster state used in MBQC~\cite{claes2023tailored} (See Appendix~\ref{app:SE_methods_info}.\ref{supp:teleportation_based_syndrome_extraction} for more details).
While typically associated with MBQC, it can also be implemented in a circuit-based framework where gates are directly applied to qubits. 
Under our error model with noise only on entangling gates, this construction yields short lifecycles of length four for all physical qubits in the bulk.
Among the SSR-based approaches considered, it achieves the shortest lifecycles without experimental operations beyond SSR, though at the cost of additional qubits. 
These features make it well-suited for neutral-atom architectures and particularly effective in loss-dominant regimes, as discussed in Sec.~\ref{sec:comparing_different_SE}.

\subsection{\textit{Mid-circuit erasure conversion SE approaches}}
\label{subsection:direct_conversion_SE}

\begin{figure}[h]
  \centering
  \includegraphics[width=\columnwidth]{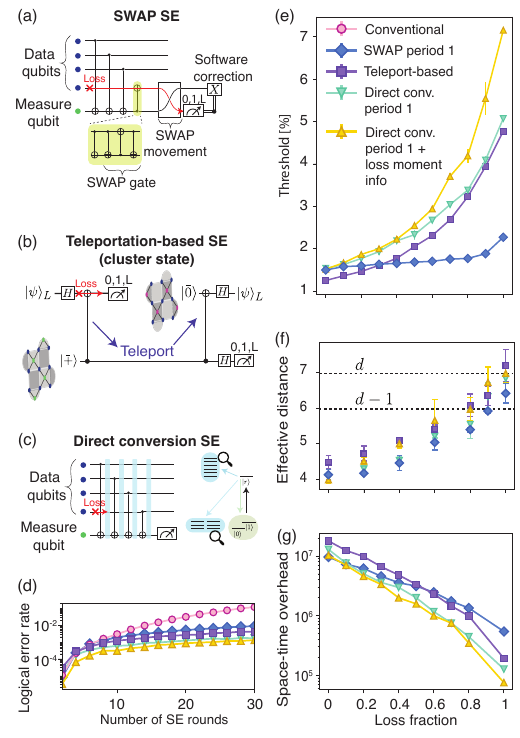}
  \caption{Loss-detecting SE methods. 
    (a)~Modified SWAP SE: data and measure qubits are swapped each round to detect loss via SSR; 
    the SWAP’s CNOT decomposition cancels with existing parity-extraction gates, with the remaining CNOT replaced by classical feedforward, so no additional entangling gates are required.
    (b)~Teleportation-based SE: logical qubits are teleported to fresh blocks prepared in alternating bases, detecting loss via SSR and shortening lifecycles, at the cost of extra qubits. 
    (c)~Direct conversion SE: mid-circuit measurement and replacement convert loss to erasure using additional hardware~\cite{wu2022erasure}; detection/replacement frequency is varied in simulations.
    (d)~Logical error rates vs.\ SE rounds ($d=7$, $p=1\%$) using the delayed-erasure decoder. Comparable performance is observed across SE methods in regimes with short lifecycles. 
    (e)~Error thresholds vs.\ loss fraction, showcasing the improvement with loss for all SE methods. 
    (f)~Effective distance vs.\ loss fraction ($d=7$), improving with loss for all SE methods.
    (g)~Space-time overhead to reach $P_L=10^{-12}$ at $p=0.5\%$ vs.\ loss fraction. 
    Legends for (d-g) appear in (e).}
  \label{fig:syndrome_extraction_methods_memory}
\end{figure}

Mid-circuit erasure conversion offers an alternative to SSR-based detection by converting physical qubit loss into flagged erasure events during circuit execution (See Fig.~\ref{fig:syndrome_extraction_methods_memory}(c).)
This has been demonstrated in alkaline-earth-like atoms~\cite{wu2022erasure,ma2023high,sahay2023high,scholl2023erasure} and superconducting qubits~\cite{kubica2023erasure,gu2023fault,chang2024surface,miao2023overcoming,levine2024demonstrating,de2025mid,chou2023demonstrating, koottandavida2024erasure, mehta2025bias}, using auxiliary fluorescence imaging, auxiliary qubits, or leakage monitors, followed by reset or transport. 
The resulting erasure is often delayed and requires active intervention, such as qubit replacement or reset, imposing experimental overhead in real-time implementations.

We refer to this as the direct conversion SE method, and explore multiple detection-replacement schedules:
(i) detection and replacement after each gate layer or SE round (periods~0.25,1), and
(ii) detection after each gate but replacement only at the end of the round (period~1~+~loss moment information).
Frequent detection improves decoder performance; frequent replacement reduces logical error by halting loss propagation, but might extend circuit duration due to feed-forward control. 
This flexibility is a practical advantage of erasure conversion.

\subsection{\textit{Comparison of SE methods}}
\label{sec:comparing_different_SE}

We now compare the performance of these approaches in the context of a surface code memory, using circuit-level simulations with varying loss fractions \mbox{$L = {p_{\text{loss}}} / ({p_{\text{loss}} + p_{\text{Pauli}}})$}, where $p_{\text{loss}}$ and $p_{\text{Pauli}}$ are physical loss and Pauli error probabilities, respectively. 
We identify the optimal SE approach at different loss fractions and Pauli error biases (see Appendix~Section~\ref{app:interplay_loss_erasure_biased_errors}
for detailed results with bias).

Figure~\ref{fig:syndrome_extraction_methods_memory}(d) compares the logical error rates of conventional SE without loss detection, approaches with loss detection—such as SWAP SE, teleportation-based SE, and direct conversion SE—and direct conversion SE with perfect loss information,
in a single logical memory experiment at an experimentally-motivated loss fraction of 0.5, comparable to that observed in recent neutral atom experiments~\cite{evered2023high, ma2023high, quinn2024high,bluvstein2025architectural}.
For a small number of SE rounds ($\leq 10$), all protocols behave similarly. However, as the number of SE rounds increases, loss detection and the replacement of lost qubits substantially enhance performance, providing a convenient approach to achieving high circuit depths without additional experimental gate overhead.

Figure~\ref{fig:syndrome_extraction_methods_memory}(e) shows the threshold of each SE method as a function of the loss fraction.
We determine the threshold by calculating the logical error rate for different code distances \( d \), using a noiseless initialization, followed by \( d-1 \) rounds of noisy stabilizer checks and a final noiseless transversal measurement.  
We find that the thresholds of all methods improve with increasing loss fraction. 
Notably, methods with shorter lifecycles, such as teleportation-based SE, benefit more from increasing loss fraction.
In the next Section~\ref{subsec:linking_performance_to_key_metrics}, we provide further insight into the thresholds of each SE method by linking it to simple characterizations such as qubit lifecycles.

Figure~\ref{fig:syndrome_extraction_methods_memory}(f)
presents the effective code distance of each SE method as a function of the loss fraction, for distance $d=7$. The effective distance $d_e$ refers to the number of errors required to cause a logical failure, as the logical error rate scales as $(p/p_{\text{th}})^{d_e}$ far below the threshold $p_{\text{th}}$, where $p$ is the physical error rate. For Pauli noise, $d_e=(d+1)/2$, and for erasure noise $d_e=d$. 
To determine $d_e$ for each loss fraction, we fit the logical error rate data far below the threshold to the function $\alpha p^{\beta}$, where $\alpha$ and $\beta$ are fitting parameters (see Fig.~\ref{fig_supp:effective_distance}).  
All SE methods, while utilizing the delayed-erasure decoder, experience increased effective distance with an increased loss fraction. 
For loss errors only, teleportation-based SE and direct conversion SE methods (periods 0.25 and 1) achieve the optimal effective distance of \mbox{$d_e\approx d$}, while SWAP SE achieves \mbox{$d_e\approx d-1$}, likely due to longer lifecycles. 
Additional threshold and effective distance results under an alternative error model from Ref.~\cite{yu2024tracking} are presented in Table~\ref{table:syndrome_comparison}.

To account for the qubit overhead required in different SE methods, we evaluate the space-time overhead for performing $d$ SE rounds of each method in Figure~\ref{fig:syndrome_extraction_methods_memory}(g). 
We define \emph{space-time overhead} as the product of the total number of physical qubits and the circuit depth required for the computation.
This metric provides a platform-independent measure of logical performance. However, it does not include experimental considerations such as real-time qubit transport, additional laser resources, or mid-circuit measurement overhead. These factors can significantly affect implementation and are discussed qualitatively in Sections~\ref{subsection:SWAP_SE}-\ref{subsection:direct_conversion_SE}.
For a given physical error rate of $0.5\%$, we determine the required code distance to achieve a logical error rate of $10^{-12}$ by fitting the logical error as a function of distance for $d = 3, 5, 7, 9$. 
We then present the space-time overhead of each approach as a function of the loss fraction, highlighting the overhead reduction as the loss fraction increases for all SE approaches.
Our results highlight that although teleportation-based SE requires more qubits, its better error suppression when the loss fraction is high leads to a more favorable space-time volume compared to SWAP SE (see space-time overhead for each SE method in Table~\ref{table:syndrome_comparison}).

At the experimentally relevant loss fraction of $\sim 0.5$, seen in multiple different hardware systems such as neutral atoms and ions ~\cite{evered2023high, ma2023high,quinn2024high,bluvstein2025architectural}, the thresholds of the different methods range from 1.5\% to 2.5\%, and all SE methods achieve similar effective distances. 
However, their experimental feasibility differs. 
Direct conversion SE requires erasure conversion with mid-circuit detection and real-time replacement of lost qubits, often demanding extra lasers and fast feedback. By contrast, SSR-based approaches such as SWAP or teleportation SE avoid these requirements. 
Thus, while all methods benefit from accurate loss detection, the optimal choice depends strongly on hardware-specific capabilities and constraints.

Another important aspect of neutral atom qubits is their intrinsic bias, as $Z$-type Pauli errors are much more common than $X$-type Pauli errors~\cite{evered2023high}. 
In Appendix Section~\ref{app:interplay_loss_erasure_biased_errors}, we investigate the interplay between biased errors and delayed erasure errors across multiple SE methods, employing the XZZX surface code~\cite{bonilla2021xzzx} and the XZZX cluster state~\cite{claes2023tailored} for teleportation-based SE. 
Our analysis considers scenarios both with and without bias-preserving gates, as well as the presence of biased erasure noise. Bias-preserving gates, such as native $CX$ gates, preserve the bias of the errors, in contrast to $CX$ gates decomposed into $CZ$ and $H$ gates. Biased erasure noise occurs when a qubit exits the computational subspace exclusively from the state $\ket{1}$. As a result, the replaced qubit follows a biased error channel instead of a depolarizing error channel.
Additionally, we investigate scenarios where Pauli errors are biased, examining the two-dimensional space defined by the bias ratio and the loss fraction. Thresholds are calculated at each point in this space for various SE methods (see Figures~\ref{fig_supp:erasure_bias_plots}, \ref{fig_supp:loss_bias_plots_memory} in the Appendix).
A key observation emerges across all SE methods: increasing the loss fraction has a significantly greater impact on thresholds than increasing the bias ratio, even when using bias-preserving gates and biased-erasure.
This difference in impact between the loss fraction and the bias ratio becomes even more pronounced in the absence of bias-preserving gates.

\begin{figure}[t]
  \centering
  \includegraphics[width=\columnwidth]{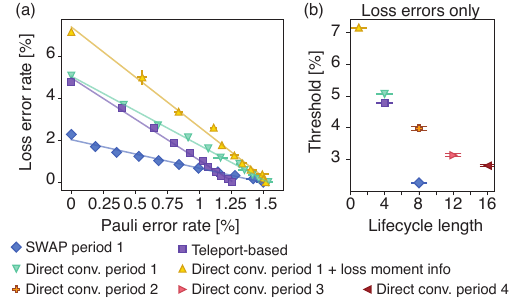}
  \caption{Linking thresholds to key metrics. (a) Thresholds for different SE methods for different loss and Pauli error rates. 
  The curves are linear fits to numerical finite-size data, with the region below each curve representing the correctable region. 
  (b) Thresholds as a function of lifecycle length for various SE methods, in the loss error only limit.
  }  
  \label{fig:Phase_diagram}
\end{figure}

\subsection{\textit{Predicting performance by error counting} \label{subsec:linking_performance_to_key_metrics}}

We now present a unified model for analyzing SE methods by linking performance to simple characterizations such as lifecycle length and the number of entangling gates.
Figure~\ref{fig:Phase_diagram}(a) shows the thresholds in the parameter space of loss and Pauli error rates for various SE methods. 
Thresholds are plotted in terms of the loss error rate and Pauli error rate, with the region below each curve representing the range of correctable errors.
We numerically determine a good fit to a linear model (solid lines) based on finite-size data, given by $ p_{\text{loss}} = p_{\text{loss,th}} - (p_{\text{loss,th}} /p_{\text{Pauli,th}}) \cdot p_{\text{Pauli}}$,
where \(p_{\text{loss,th}}\) and \(p_{\text{Pauli,th}}\) are the respective thresholds for loss and Pauli errors only.
The linear behavior suggests that the threshold depends on the loss fraction $L$ according to the relationship $p_{\text{threshold}} = p_{\text{loss,th}} p_{\text{Pauli,th}} /(L(p_{\text{Pauli,th}} - p_{\text{loss,th}}) + p_{\text{loss,th}})$, and is qualitatively similar to the behavior observed in Refs.~\cite{barrett2010fault, stace2010error, stace2009thresholds, perrin2024quantum}.

The curve intersections with the axes provide key insights. The y-axis intersection, representing the threshold in the absence of Pauli errors, is related to the lifecycle length. 
For the direct conversion SE period 1 with perfect loss information, 
the threshold is \(p_{\text{loss,th}} \approx 7.2 \%\). 
Increasing the lifecycle reduces the threshold, which decays with the lifecycle length, as shown in Fig.~\ref{fig:Phase_diagram}(b).
Notably, the SWAP SE, with an average lifecycle length of 8, has a lower threshold of \(p_{\text{loss,th}} \approx 2.3\%\), 
deviating from the heuristic. 
This lower performance is attributed to additional factors unique to SWAP SE, such as the loss of stabilizer information when the swapped in data qubit is lost and the fact that measurement qubit loss is only detected in the following round, after it propagates as a data qubit loss, as discussed in Section~\ref{subsection:SWAP_SE}.

The x-axis intersection, representing Pauli error thresholds, correlates with the number of entangling gates, which impacts the number of Pauli errors in the final state. For example, over $d$ SE rounds, teleportation-based SE uses $1.5\times$ more entangling gates than SWAP SE and direct conversion SE approaches, resulting in a proportional decrease in the Pauli error threshold, as similarly observed in Ref.~\cite{sahay2023high}.

\section{The effect of Loss Errors in deep Logical Algorithms design}
\label{sec:loss_random_algs}
\begin{figure}[t]
  \centering
  \includegraphics[width=\columnwidth]{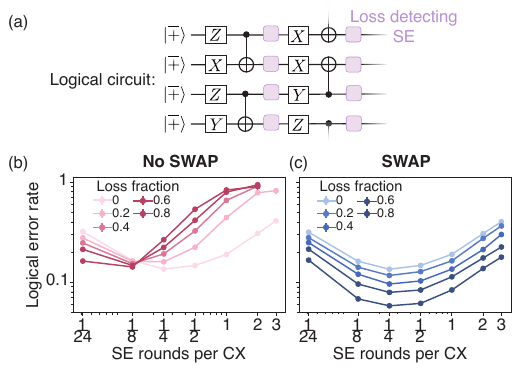}
  \caption{Deep logical circuits with qubit loss. (a) A deep Clifford logical algorithm consisting of random logical single-qubit and transversal $CX$ gate layers, with periodic SE rounds at varying frequencies. (b, c) Circuit-level simulation results showing the logical error rate as a function of the number of SE rounds per transversal gate layer, for different loss fractions.
  ($p=1\%$, $d=5$, 24 layers). 
  The SWAP SE method (c) effectively mitigates loss errors, restoring the optimal SE frequency observed in Pauli-dominated scenarios. In contrast, conventional SE (b) exhibits varying error correction regimes, where loss can either improve or degrade performance depending on the lifecycle length. 
  }
  \label{fig:Random_Logical_algorithms}
\end{figure}

Loss errors in logical algorithms can have significantly different effects compared to standard memory benchmarks. 
We now study the effects of loss errors on the QEC design of multi-qubit deep logical algorithms.
Specifically, we analyze how physical loss errors detection influence the optimal frequency of SE rounds between transversal gates. Our analysis reveals that SWAP-based SE achieves comparable optimal SE frequencies in both the presence and absence of loss.

We focus on the effects of loss in random Clifford logical algorithms (Fig.~\ref{fig:Random_Logical_algorithms}(a)), where multiple logical qubits interact through transversal gates interspersed with periodic SE rounds at tunable frequencies. Circuit-level simulations are performed for logical circuits comprising 24 layers of logical $CX$ and single-qubit logical gates ($X,Y,Z$), with SE rounds applied at varying intervals, as in Ref.~\cite{cain2024correlated} (for more details see Appendix~\ref{app:logical_algorithms}). 
Two scenarios are analyzed: conventional SE without explicit loss detection, which detects only measurement qubit loss in each SE round (Fig.~\ref{fig:Random_Logical_algorithms}(b)), and SWAP-based SE (Fig.~\ref{fig:Random_Logical_algorithms}(c)). 
The x-axis represents the number of SE rounds per gate layer, ranging from no SE~($n_r=1/24$) to multiple SE rounds after every gate layer~($n_r=3$).

In the absence of SWAP operations (but still leveraging SSR detection and delayed-erasure decoding), we observe several distinct error correction regimes. For a small number of SE rounds per gate, loss improves the logical error rate due to short qubit lifecycles. However, as the number of SE rounds per gate increases, performance degrades due to the cumulative effects of loss over longer lifecycles. 
By contrast, incorporating SWAP-based SE, loss consistently enhances error correction performance. Loss detection via SWAP stabilizer checks restores the previous optimal SE rounds per $CX$ observed in Ref.~\cite{cain2024correlated}. Consequently, when employing SWAP SE and SSR, the presence of loss does not alter the heuristic conclusions made for logical algorithms dominated by Pauli errors and can, in fact, improve overall performance.

\section{Native loss detection from logical teleportation}
\label{sec:logical_algs_with_teleportation}
\begin{figure}[t]
  \centering
  \includegraphics[width=\columnwidth]{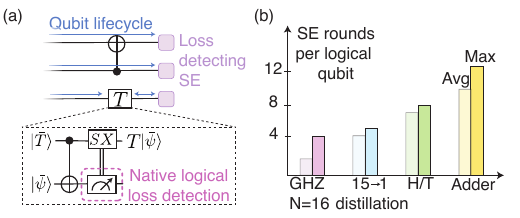}
  \caption{Native loss detection in logical algorithms. 
  (a)~Logical qubit teleportation during gate teleportation, shortening lifecycles and naturally detecting loss without loss-detecting SE rounds. 
  (b)~SE rounds per logical qubit before logical measurement for various key algorithmic subroutines. Each subroutine uses frequent gate teleportation, keeping lifecycles short and detecting loss without additional experimental overhead. 
\label{fig:Logical_procedures_teleportation}
  }
\end{figure}

While we have techniques for handling algorithmic structures with long qubit lifecycles, we now observe that in many cases, qubit lifecycles are short in realistic logical algorithms. 
In particular, teleportation is a powerful technique for loss detection and correction that avoids additional overhead, and it naturally emerges within logical algorithms through gate teleportation (see Fig.~\ref{fig:Logical_procedures_teleportation}(a)). 
Specifically, the SWAP-teleported gates exchange the logical data qubit with a teleported logical qubit that implements the desired gate, detecting loss using SSR and terminating the qubit lifecycles without the need for additional SE rounds.

This approach is further enhanced by correlated decoding and algorithmic fault tolerance, which capitalize on the use of transversal gates and teleportation in universal quantum computation~\cite{cain2024correlated,zhou2024algorithmic}. These ensure that lifecycles are inherently short in many logical algorithms, simplifying loss management while enabling regular qubit replacement.

A key observation is that a wide range of known logical subroutines naturally employ teleportation, inherently keeping qubit lifecycles short.
Fig.~\ref{fig:Logical_procedures_teleportation}(b) illustrates the average number of SE rounds per physical qubit, from initialization to measurement, across various essential subroutines. The results show that most algorithms have relatively brief lifecycles. Detailed descriptions of these algorithms are provided in Appendix~\ref{app:logical_algorithms}. For this analysis, we conservatively assume one SE round per logical operation. However, as shown in Fig.~\ref{fig:Random_Logical_algorithms}(c), this number can be further reduced. Additionally, Fig.~\ref{fig:syndrome_extraction_methods_memory}(d) and Fig.~\ref{fig_supp:LER_num_cycles_all_L} reveal that, at an experimentally-motivated loss fraction of 0.5, for fewer than 10 SE rounds, performance remains nearly unchanged without any active loss correction added to the SE rounds, relying solely on SSR and delayed-erasure decoding.
These findings suggest that while memory benchmarks provide useful performance metrics, transversal logical algorithms are inherently well-equipped to manage loss, with minimal to no loss detecting SE. 
Leveraging SSR and delayed-erasure decoding can significantly enhance the performance of the logical algorithm without incurring additional experimental complexity.

\begin{figure}[t]
  \centering
  \includegraphics[width=\columnwidth]{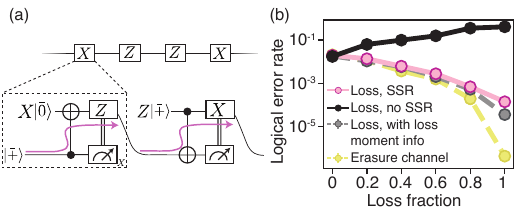}
  \caption{Deep logical algorithms with teleported gates.
  (a)~We study a deep circuit consisting of teleported $X$ and $Z$ logical gates, similar in structure to small-angle synthesis algorithms~\cite{kitaev1997quantum, kitaev2002classical}.
  (b)~Circuit-level simulation results for 11 layers of random teleported $Z$ and $X$ logical gates, with all logical qubits initialized in the presence of noise (physical error rate $p=1\%$ and $d=7$). 
    The delayed-erasure decoder (pink) substantially outperforms an MLE decoder that does not account for loss (black) and improves with increasing loss fraction.
    It also matches the performance of a decoder with perfect loss information (gray, dashed), and for loss fractions less than 1, approaches that of the erasure channel (khaki, dashed).
    The erasure channel performance is achieved using direct conversion SE with period 0.25, corresponding to loss detection and qubit replacement after every gate, thus providing a lower bound.
  }
  \label{fig:Logical_telep_simulations}
\end{figure}

As a proof of concept, we consider the teleportation-based logical circuit shown in Fig.~\ref{fig:Logical_telep_simulations}(a), which mirrors the structure of the small-angle synthesis algorithm~\cite{kitaev1997quantum, kitaev2002classical}. This algorithm constructs small-angle rotations through sequences of $H$ and $T$ gates, with the latter often realized via teleportation. Consequently, loss detection and correction are integrated into the process of executing the logical algorithm.
The logical error rates for the circuit in Fig.~\ref{fig:Logical_telep_simulations}(a) are plotted in Fig.~\ref{fig:Logical_telep_simulations}(b) as a function of the loss fraction using various decoders: a delayed-erasure decoder leveraging SSR, a regular MLE decoder, and an MLE decoder with perfect loss time-location information.
The delayed-erasure decoder demonstrates significant performance improvements, with algorithmic logical error rates rapidly decreasing as the loss fraction increases, achieving the performance of the decoder with perfect loss information.
Furthermore, for loss fractions less than 1, it achieves the lower bound set by the erasure channel—corresponding to loss detection and qubit replacement after every gate.
Notably, this circuit is similar to the teleportation-based SE method investigated here in the memory setting, both utilizing SSR and replacing atoms frequently through teleportation without additional overhead.
These results highlight the advantages of exploiting native loss detection within teleportation gadgets in logical circuits while also utilizing loss information to improve decoding.
Finally, while our simulations focus on Clifford gate teleportation, they are expected to extend to non-Clifford gates, as the underlying loss detection principles remain consistent.

\section{Conclusion}
\label{sec:conclusion}

These results provide a general framework for managing qubit loss or leakage errors across diverse quantum computing platforms, primarily neutral atoms but also including superconducting qubits and trapped ions, 
and highlight the impact of loss detection and bias on logical algorithm performance.
Our analysis across a wide range of loss fractions and detection strategies reveals how loss detection and decoding can be leveraged in realistic regimes, and was already employed in the recent experimental studies of architectural mechanisms for universal fault-tolerant quantum computing using neutral 
atom arrays~\cite{bluvstein2025architectural}.

Central to our approach is the use of the delayed-erasure decoder, which leverages loss information to approximate an MLE decoding solution, substantially improving the logical error rate.
Using our decoder, we examine how the algorithmic structure influences the optimal strategy for detecting and correcting loss.
For high-depth circuits involving a large number of gate layers prior to logical measurement, the performance depends on both loss decoding and frequent loss detection and replacement.
Using a surface code logical memory, 
we explore different SE methods under various loss fractions.  Our findings indicate that, with appropriate decoding strategies, all methods enhance performance with increasing loss fraction. 
Notably, teleportation-based SE is a promising candidate for neutral atom quantum computing, as it leverages SSR to achieve comparably high thresholds for high loss fractions, albeit with an additional space overhead.
Determining the optimal SE method for a given system will depend on experimental validation and the specific noise characteristics of the hardware.

By applying these results to multi-qubit deep logical algorithms, several key insights emerge. Loss errors, when managed using loss detection operations and SSR detection, are fully compatible with correlated decoding, allowing for an optimal number of approximately four logical operations per SE round using the delayed-erasure MLE decoder. This framework ensures effective error correction even in the presence of high loss rates.

By considering key subroutines involving extensive logical teleportation, we find that the use of transversal gates and correlated decoding keeps the number of SE rounds sufficiently small, such that loss is natively detected and managed by the gate teleportation intrinsic to universal processing. 
Examining a toy model of a small-angle synthesis algorithm, we observe that loss errors, decoded using our delayed erasure decoder, significantly enhance performance compared to Pauli channels solely through logical teleportation.
As such, while the logical memory benchmark provides valuable insights into various strategies for loss management and detection to enhance performance, behavior can radically change when realizing logical algorithms.

This work opens up several new research directions. 
A major priority is reducing the decoding runtime, critical for scaling to more complex and larger algorithms.
While the present work mostly employs an MLE inner decoder, its limited scalability for large-scale logical algorithms underscores the need for further refinement, possibly through alternative strategies, such as matching decoders~\cite{google2023sparse, wu2023fusion, google2024quantum} or machine learning-based decoders~\cite{google2024learning, google2024quantum}.

An intriguing avenue for future exploration involves optimizing loss detection in specific algorithmic subroutines, tailoring the loss detection frequency and SE method to align with the algorithmic gadget and system characteristics.
Additionally, the delayed-erasure decoder can be naturally applied to the decoding of transversal non-Clifford circuits, as it is inherently compatible with and effective for correcting Clifford errors propagating through the circuit. 

Finally, although our simulations focus on the surface code, the core techniques presented in this work are broadly applicable to a wide range of QEC codes and logical algorithms. 
In particular, our delayed-erasure decoder is agnostic to the underlying code and can be applied to any stabilizer code with a suitable decoding backend. 
Moreover, several loss-handling strategies explored here, such as erasure conversion and teleportation-based syndrome extraction, are compatible with a wide class of codes, including high-rate qLDPC codes~\cite{tillich_2014, Panteleev2021degeneratequantum, leverrier2022quantumtannercodes, pantaleev_linear_distance, pantaleev_kalachev, breuckmann_ldpc,xu2024constant,xu2024fast, ataides2025constant}.

\textit{Note added}. Since the completion of our work, we became aware of several related studies examining the effect of qubit loss on a single logical memory: teleportation-SE~\cite{yu2024tracking}, loss detection units~\cite{perrin2024quantum}, and SWAP-SE~\cite{yu2025SWAP}. 
While addressing similar challenges, these works differ from ours in focus and methodology. Notably, improved thresholds have been recently reported in updated versions of Refs.\cite{yu2024tracking, yu2025SWAP}, based on the use of decoding strategies similar to the delayed erasure decoder described in this work. These contributions also utilize error models slightly different from those presented here. 
For completeness, our Appendices includes results obtained using the error models utilized in Refs. ~\cite{yu2024tracking, yu2025SWAP}.

\let\oldaddcontentsline\addcontentsline
\renewcommand{\addcontentsline}[3]{} 
\section*{Acknowledgments}
We thank Sasha Geim, Alex Kubica, Jeff Thompson, Liang Jiang, Isaac Chuang, Nazli Ugur Koyluoglu, Luke Stewart, Marcin Kalinowski, Edita Bytyqi, David C. Spierings, Katie Chang, Casey Duckering, Shengtao Wang, Chen Zhao for insightful
discussions and  Susanne Yelin for her guidance and valuable discussions.
G.B. acknowledges support from the MIT Patrons of Physics Fellows Society.
D.B. acknowledges support from the NSF GRFP (grant DGE1745303) and the Fannie and John Hertz Foundation.
We acknowledge financial support from IARPA and the Army Research Office, under the Entangled Logical Qubits program (Cooperative Agreement Number W911NF-23-2-0219), the DARPA IMPAQT program (HR0011-23-3-0012), and the DARPA MeasQuIT program (HR0011-24-9-0359), QuEra Computing Inc (award number A57912), the Center for Ultracold Atoms (an NSF Frontier Center), and the National Science Foundation (award number PHY-2012023).

\appendix

\section{Syndrome of Loss Errors}
\label{app:syndrome_of_loss_errors}

This section explains how heralded loss alters the stabilizer measurement structure and modifies the decoding hypergraph in QEC codes. 
Each loss mechanism induces a modified circuit, which alters the set of stabilizer detectors and introduces new correlated error configurations. 
The resulting decoding hypergraph consists of a set of detectors (vertices) and hyperedges, where each hyperedge represents a possible error event, connects a subset of detectors it flips, and is assigned a corresponding probability. 
This hypergraph serves as the input to the decoder.
Given a heralded loss at measurement (e.g., using SSR or erasure conversion), we reconstruct the lifecycle of the affected qubit (defined as the sequence of operations from initialization to measurement) and enumerate its potential loss locations. 
Each potential loss point defines a distinct “loss circuit,” which produces a different set of hyperedges in the decoding hypergraph.

As discussed in the main text, loss requires two decoder-level adjustments:
(1) modifying detectors to remove dependence on the lost measurement (via superchecks), and 
(2) accounting for error edges arising from truncated circuits after loss.
Below, we elaborate on each adjustment and provide a specific example for clarity.

First, in each loss circuit, the measurement of the lost qubit is set to a random result. This results in assigning a probability of $0.5$ to the relevant edge, which corresponds to a weight of $0$ in the decoding hypergraph. This process generates a supercheck operator, defined as the product of neighboring checks, effectively eliminating the lost qubit's contribution \cite{stace2009thresholds}. 
Since the loss is detected in all loss circuits for the potential loss locations, the supercheck operator appears in all decoding hypergraphs of the qubit's lifecycle. Consequently, the supercheck is incorporated into the final decoding hypergraph of the lifecycle.

Second, when an atomic qubit is lost, subsequent gates involving this qubit act trivially. Therefore, each loss circuit omits certain gates due to the lost qubit. As a result, the decoding hypergraph for the circuit includes new edges that correspond to these missing gates, capturing the errors associated with the specific loss under consideration. These edges appear in the final decoding hypergraph of the lifecycle, with probabilities reflecting the likelihood of the associated events.

\begin{figure}[h]
  \centering
  \includegraphics[width=\columnwidth]{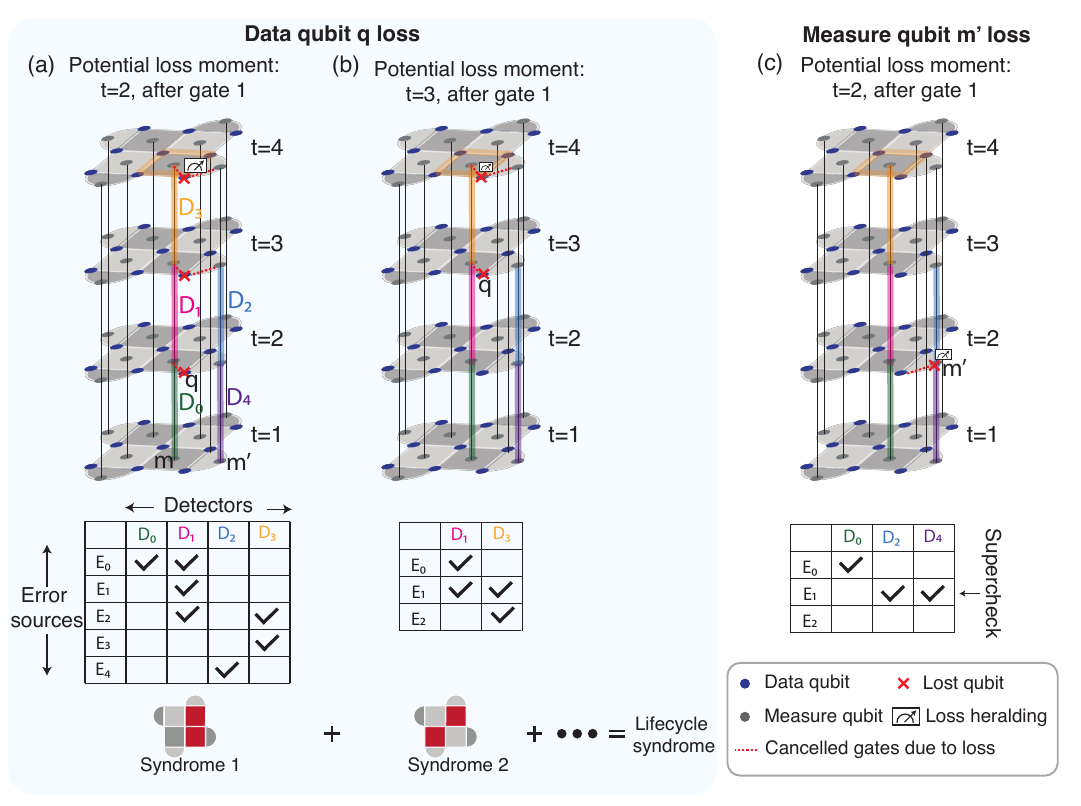}
  \caption{
Example of detector activations generated by different loss events in a \(d = 3\) surface code over four time steps (\(t = 1\)–4). 
Panels (a) and (b) show two potential loss locations for the same data qubit \(q\), in round \(t = 2\) and \(t = 3\), respectively, each after the first gate. 
These correspond to different loss circuits within the same lifecycle and are combined into a single decoding hypergraph for that lifecycle (labeled ``lifecycle syndrome''). 
Panel (c) shows a loss of a measure qubit \(m'\) in round \(t = 2\), which belongs to a separate lifecycle and contributes a separate decoding hypergraph. 
Detectors \(D_0\)–\(D_4\) are highlighted in color. 
Each row (\(E_i\)) in the decoding hypergraph tables corresponds to one error mechanism (hyperedge), i.e., a correlated set of detector activations triggered by a single loss event. 
Superchecks that are activated in all loss circuits of a given lifecycle also appear in the final decoding hypergraph, such as the shared activation of \(D_3\) in both (a) and (b).
}
  \label{fig:loss_syndrome_example}
\end{figure}

Figure \ref{fig:loss_syndrome_example} illustrates a specific example using a \(d=3\) surface code over four time steps: an initialization round ($t=1$), two syndrome extraction (SE) rounds ($t=2,3$), and a final transversal measurement ($t=4$). 
In this example, we present the syndrome patterns caused by losses at different locations and times. 
The upper panels show the circuits with canceled gates and heralding, while the lower panels display the decoding hypergraph, listing which detectors are activated and their correlations. 
This example uses a conventional SE scheme with measure qubits read out every round and data qubits only at the end, but our decoding approach applies to any SE method, as discussed in the main text.

Detectors \(D_0\)-\(D_4\) correspond to parity checks between stabilizer outcomes across rounds:
\(D_0\) and \(D_1\) compare the same stabilizer across \(t=1\!\to\!2\) and \(t=2\!\to\!3\) respectively,  
\(D_2\) monitors the neighboring stabilizer of \(q\) across \(t=2\!\to\!3\), 
\(D_3\) is the final plaquette stabilizer containing \(q\) given by the final transversal measurement and compared with previous round,  
and \(D_4\) compares measure qubit \(m’\) across \(t=1\!\to\!2\).

Each row in the decoding hypergraph (Figure~\ref{fig:loss_syndrome_example}) corresponds to a distinct detector activation pattern produced by propagating a specific loss through the truncated circuit. 
For example, when data qubit $q$ is lost at $t=2$, all subsequent CZ gates on $q$ are canceled and the final readout is randomized. 
The missing CZs make stabilizers involving $q$ inconsistent across rounds, producing time-like detector flips such as $\{D_0,D_1\}$ or $\{D_1\}$. 
In some branches, the loss heralding itself introduces a plaquette mismatch ($D_3$), or the neighboring stabilizer of $q$ registers a time-like inconsistency ($D_2$). 
Thus the multiple rows for a single loss represent different deterministic ways a single heralded loss can violate stabilizer checks, depending on timing and Pauli by-product branches.
Now we explain each row explicitly.

\textbf{(a) Loss of data qubit \(q\) at \(t=2\).}  
All subsequent gates on \(q\) are canceled, and the final measurement is probabilistic. 
The missing CZs disrupt stabilizers involving \(q\). 
The decoding hypergraph contains five rows:  
\(\{D_0,D_1\}\) (mismatches in both \(t=1\!\to\!2\) and \(t=2\!\to\!3\)),
\(\{D_1\}\) (only \(t=2\!\to\!3\)),   
\(\{D_1,D_3\}\) (round \(t=2\!\to\!3\) plus the plaquette check containing \(q\)),  
\(\{D_2\}\) (neighbor stabilizer at \(t=2\!\to\!3\)), and   
\(\{D_3\}\) (plaquette \(t=3\!\to\!4\) inconsistency).

\textbf{(b) Loss of data qubit \(q\) at \(t=3\).}  
Since stabilizers remain intact through \(t=2\), only the final checks are affected:  
\(\{D_1\}\) (round \(t=2\!\to\!3\)),  
\(\{D_1,D_3\}\) (round \(t=2\!\to\!3\) plus plaquette \(t=3\!\to\!4\)), and  
\(\{D_3\}\) (plaquette \(t=3\!\to\!4\) inconsistency).

Since both (a) and (b) correspond to different loss locations within the same data qubit lifecycle, any detectors that are activated in all loss circuits become part of the supercheck structure for that lifecycle.
In this case, detector \(D_3\), which corresponds to the final plaquette stabilizer involving \(q\), is activated in both loss circuits and thus appears in both decoding hypergraphs. 
As a result, \(D_3\) is retained in the final decoding hypergraph for the lifecycle and serves as a supercheck.

\textbf{(c) Loss of measure qubit \(m’\) at \(t=2\).}  
The round-2 measurement of \(m’\) is missing, invalidating detectors \(D_2\) and \(D_4\). 
Their product forms a valid supercheck, \(D_2 D_4 = M_{m’,t=1}M_{m’,t=3}\), correlating rounds 1 and 3 directly.
Additionally, the canceled CZ introduces an error on the paired data qubit, activating \(D_0\).

These examples show how a single loss modifies the circuit and produces specific correlated detector patterns.

\section{Delayed-Erasure Decoder}
\label{appendix:delayed_erasure_decoder}

This section presents our decoding strategy for handling heralded loss in fault-tolerant circuits.
Loss acts in a much more complex fashion compared to normal Pauli errors, and therefore requires more care in terms of their formulation.
Here, we show how to formulate the full MLE problem, and then construct a scalable decoder that approximates this solution.
\subsection{Exact MLE Decoding in the Presence of Loss}
\label{appendix:mle_decoder}

This section formulates the MLE problem in the presence of heralded qubit loss. 
While exact MLE decoding is computationally challenging, it serves as a theoretical benchmark that guides the development of efficient approximate methods.

We generalize the framework of Ref.~\cite{cain2024correlated} by incorporating both Pauli and loss errors. 
Let \( \vec{E} = (E_1, E_2, \ldots) \) represent binary variables for Pauli error events, and \( \vec{L} = (L_1, L_2, \ldots) \) represent binary variables for loss events. 
We assume Pauli and loss errors occur independently across circuit locations, although correlated models can be incorporated.

Measurement outcomes provide two kinds of information:
(1) Flags \( \vec{F} \), indicating when qubit loss is detected;
and (2) Detectors \( \vec{D} \), constructed as products of stabilizer measurement outcomes that are expected to be \( +1 \) in the absence of error. 
Because loss renders some measurements invalid, the set of active detectors must be dynamically redefined based on the observed flags.

The most likely error problem seeks to find the assignment of Pauli errors $\vec{E}$ and loss errors $\vec{L}$ that maximizes the following quantity:
\begin{align}
P(\vec{E}, \vec{L}|\vec{F}, \vec{D}),
\end{align}
Namely, given the observed flags, and the resulting detectors that we define conditional on seeing those flags, what is the most likely error.

We can rewrite this expression using Bayes' rule
\begin{align}
P(\vec{E}, \vec{L}|\vec{F}, \vec{D})&=\frac{P(\vec{E}, \vec{L}, \vec{F}, \vec{D})}{P(\vec{F}, \vec{D})}
=\frac{P(\vec{F}, \vec{D}|\vec{E}, \vec{L})P(\vec{E}, \vec{L})}{P(\vec{F}, \vec{D})}.
\end{align}
Since we are given a fixed observation $\vec{F}$, $\vec{D}$, the denominator is fixed.
We therefore seek to maximize the product of the probability of the error configuration $P(\vec{E},\vec{L})$ and the conditional probability of observing these syndromes.
In the presence of loss, an important distinction from the usual case is that the latter factor is no longer always 0 or 1.

The error probability $P(\vec{E},\vec{L})$ can be calculated based on the log likelihood ratios of individual events.
However, since only one loss event can occur per qubit lifecycle, the effective probability of loss at location $j$ becomes $p_j \prod_{k<j}(1 - p_k)$. 
This correction is relevant when the loss probabilities vary significantly across locations.

The more challenging factor to evaluate is $P(\vec{F}, \vec{D}|\vec{E}, \vec{L})$.
First, given a loss pattern, detectors must be updated by replacing invalidated measurements with combinations of adjacent ones (e.g., superchecks). 
While the exact choice is irrelevant in full MLE decoding due to stabilizer equivalence, it can affect the performance of heuristic decoders.
After forming the detectors, we need to evaluate whether a given loss and error pattern is consistent with the observed loss flags and detectors.
The flag check is straightforward: at most one loss event should be active per qubit lifecycle.
Evaluating the detector consistency is more subtle. 
In the presence of loss, certain detector activations become probabilistic, depending on both the location and timing of the loss event. 
Given a specific loss pattern, the circuit becomes fixed, and the resulting conditional detector distribution can in principle be computed by propagating the error configuration through the modified circuit. 
However, this approach is computationally expensive and difficult to express as a simple optimization problem, motivating approximate methods.

One such approach is edge reweighting, where known correlations between loss and downstream Pauli errors are captured by modifying decoder edge weights. 
For instance, Ref.~\cite{suchara2015leakage} models leaked qubits as inducing depolarizing noise on all subsequent interactions, assigning higher error weights to gates further in time. 
While this captures leading-order effects, it neglects correlations introduced by the fact that leaked qubits persist across time steps, resulting in temporally correlated error patterns. 
Accounting for this structure can improve heuristic approximations to MLE decoding.

Ultimately, the goal is to identify practical heuristic methods that approximate MLE decoding while capturing the effects of loss. 
This can be pursued by analyzing which errors are commonly misclassified by a given weighting scheme or by benchmarking against exact MLE decoding on small instances. 
Neural network decoders offer a complementary strategy by directly learning mappings from syndrome and flag data to logical corrections, implicitly capturing both error and loss correlations~\cite{google2023suppressing}. 
However, for near-term systems where clarity and integration with standard decoders are important, improving heuristic approximations to MLE may offer greater practical benefits.

\subsection{Approximate MLE Decoding}
\label{appendix:approx-mle}

\begin{figure}[t]
  \centering
  \includegraphics[width=\columnwidth]{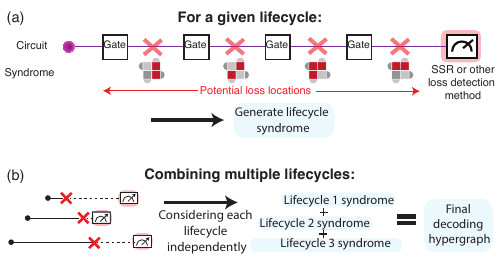}
  \caption{
  Overview of the approximate MLE decoding procedure.
  (a) For a given qubit lifecycle, we identify potential loss locations (red crosses) and simulate modified circuits for each, generating a set of decoding hypergraphs (\text{DEM}$_{ij}$).
  These include updated stabilizer syndromes (detectors) and correlated error edges resulting from the truncated loss circuits. 
  (b) When multiple qubit losses are present in a given shot, we compute the decoding hypergraph for each lifecycle independently (left), optionally include a combined loss decoding hypergraph (right), and sum with the standard Pauli decoding hypergraph to generate the final decoding hypergraph (\text{DEM}$_\mathrm{final}$), according to Eq.~\ref{eq:final_DEM}.
  This forms the input to the decoder.
  }
  \label{fig:app_aproximate_MLE_decoder}
\end{figure}

In this section, we describe our practical decoding strategy that approximates the MLE decoding introduced above. 
The goal is to capture the key features of loss-induced error correlations, particularly the way loss truncates gate sequences and alters detector structure, without solving the full combinatorial problem.

As stated in the main text, our goal is to automatically construct a decoding hypergraph based on the observed loss, which serves as the input to any general decoder for the QEC code analyzed. 
Our general approach relies on precomputing the decoding hypergraph for each possible loss location within a qubit's lifecycle.
Given a heralded loss upon measurement, we infer that the loss occurred at some earlier location in the lifecycle, and assign a probability to each possible location.
Given a specific experiment with a given loss pattern, we can sum over the relevant precomputed lifecycle decoding hypergraphs, as described below.
Figure~\ref{fig:app_aproximate_MLE_decoder} illustrates this procedure, showing how the loss locations in each lifecycle contribute to the overall decoding hypergraph for a given shot.

We now formalize our approach.
Let \( C_i \) denote the lifecycle of qubit \( i \), and let \( \{L_{ij}\} \) denote the set of possible loss locations for that qubit, each associated with a probability \( p_{ij} \). 
For each possible loss event \( L_{ij} \), we simulate the loss circuit, a modified circuit generated by removing all gates after the loss and setting the qubit’s final measurement to a uniformly random outcome. 
Using Stim’s gauge detector infrastructure~\cite{gidney2021stim}, we extract the decoding hypergraph denoted \( \text{DEM}_{ij} \), which consists of a set of hyperedges (error events) and their associated detector configurations.

Each decoding hypergraph $\text{DEM}_{ij}$ is a matrix, a collection of error probabilities $p_{n}$ and corresponding detector configurations $D_{n}$: 
\begin{equation}
    \text{DEM}_{ij} = \{ (D_{n}, p_{n}) \}, 
\end{equation}

where the columns $D_{n} = [D_{n}(0), D_{n}(1), \cdots]$ represent the list of detectors triggered by the error event in each row. 
Repeating this process for all potential loss events $L_{ij}$ within a lifecycle $C_i$ results in a set of decoding hypergraphs $\text{DEM}_{ij}$ and their associated probabilities $p_{ij}$.
This is done in a pre-processing stage.

To compute the final decoding hypergraphs for a lifecycle $C_i$ of a lost qubit $i$, we sum over all decoding hypergraphs in the lifecycle:
\begin{equation}
    \text{DEM}_i = \sum_j  \tilde p_{ij} \cdot \text{DEM}_{ij}.
\end{equation}
with normalized weights
\(\tilde p_{ij}=\Pr(L_{ij}\mid \text{flag in }C_i)\);
this is an approximation that assumes
at most one loss per lifecycle and neglects cross-lifecycle combinations.

When multiple qubit losses occur in different lifecycles, loss errors can cause correlated Clifford errors. Combining syndromes from distinct lifecycles may interfere non-linearly, resulting in syndromes that cannot be derived from independent consideration of loss events. While the optimal decoder would consider all combinations of potential loss locations for all qubits, this approach scales poorly with multiple lossy lifecycles and potential loss events, making it impractical for small code distances ($d < 7$).

To address this, we explore a decoder that evaluates each lifecycle $C_i$ independently and averages over them (\( \text{DEM}_i \)). 
Additionally, we introduce the option to include a single combination of losses across different lifecycles. 
Specifically, we consider the first potential loss locations $L_{i1}$ for all lossy lifecycles $L_i$ and generate a corresponding decoding hypergraph, $\text{DEM}_{\text{first comb}}$.

The decoding hypergraphs for individual potential loss locations $L_{ij}$ are generated in a pre-processing step, independent of any specific error model or probabilities of Pauli and loss errors. 
During real-time decoding, these hypergraphs are summed using the specific error model probabilities $\tilde p_{ij}$. 
Note that we separately account for loss errors and Pauli errors. Lossless circuits (without losses) are used to efficiently generate the decoding hypergraph for Pauli errors, denoted $\text{DEM}_{\text{Pauli}}$.

For a specific shot with a heralded loss pattern using SSR and a given error model, the final decoding hypergraph is computed by summing:
1. Lossy lifecycle decoding hypergraphs $\text{DEM}_i$,
2. The Pauli decoding hypergraph $\text{DEM}_{\text{Pauli}}$, and
3. The first potential loss combination decoding hypergraph $\text{DEM}_{\text{first comb}}$.

The final decoding hypergraph is given by:
\begin{equation}
\label{eq:final_DEM}
    \text{DEM}_{\text{final}} = \sum_i \text{DEM}_i + \text{DEM}_{\text{Pauli}} + \omega \cdot \text{DEM}_{\text{first comb}},
\end{equation}
where $\omega$ is a combination weight determined based on the analysis in Appendix Section~\ref{appendix:delayed_erasure_decoder}.\ref{appendix:approx-mle}.\ref{subsection:comb_weight}.

The summation of probabilities is calculated using the following equation:
\begin{equation}
\sum_{i=1}^l p_i \prod_{j \neq i}(1-p_j) + \sum_{i=1}^l \sum_{j=i+1}^l \sum_{k=j+1}^l p_i p_j p_k \prod_{m \neq i,j,k} (1-p_m).
\end{equation}
This ensures accuracy up to $O(p^3)$.

Once \( \text{DEM}_{\text{final}} \) is constructed for a given shot, any decoder that accepts a hypergraph-based detector error model may be used—such as MLE, MWPM, belief propagation or others, depending on the QEC code in use.

\subsubsection{Adjusting Loss Combination Weight}
\label{subsection:comb_weight}

In this section, we explore the influence of the weight of the combination, $\omega$, on the efficacy of the delayed erasure decoder, which considers both independent error events and the first combination of potential loss events. As described earlier, the decoder's decision-making process involves summing different decoding hypergraphs (see Eq.~\ref{eq:final_DEM}). To incorporate the impact of individual loss events, the first loss combination event is scaled by $\omega$ along with the standard Pauli error decoding hypergraph.

To benchmark the performance of our decoder and determine the optimal value of $\omega$, we performed circuit-level simulations on a memory logical qubit with multiple rounds of conventional syndrome extraction (SE). We varied $\omega$ between 0 and 1. For $\omega = 0$, each lifecycle's decoding hypergraph is considered independently, without including an additional decoding hypergraph for the first combination.

\begin{figure}[h]
  \centering
  \includegraphics[width=\columnwidth]{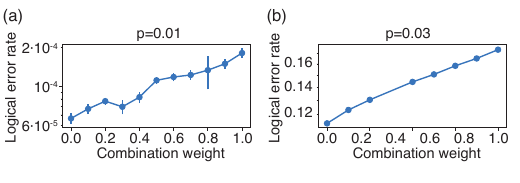}
    \caption{Logical memory simulations with $d$ rounds of the conventional SE method. Variation of the logical error rate as a function of the combination weight $\omega$ for distance 9, across different physical error rates ($p = 0.01$ and $p = 0.03$). The results indicate that nonzero values of $\omega$ deteriorate the decoder's performance.}
  \label{fig:Loss_decoder_comb_weight}
\end{figure}

The results for the conventional SE method are presented in Fig.~\ref{fig:Loss_decoder_comb_weight}. Non-zero values of $\omega$ degrade the decoder's performance. However, it is important to note that these results pertain to the conventional SE method, and different SE methods employing distinct gate sets may behave differently. For simplicity, the numerical simulations presented in this paper assume $\omega = 0$.

\section{The interplay between loss, erasure and biased errors}
\label{app:interplay_loss_erasure_biased_errors}
\subsection{The interplay between erasure and biased errors}

\begin{figure}[h]
  \centering
  \includegraphics[width=\columnwidth]{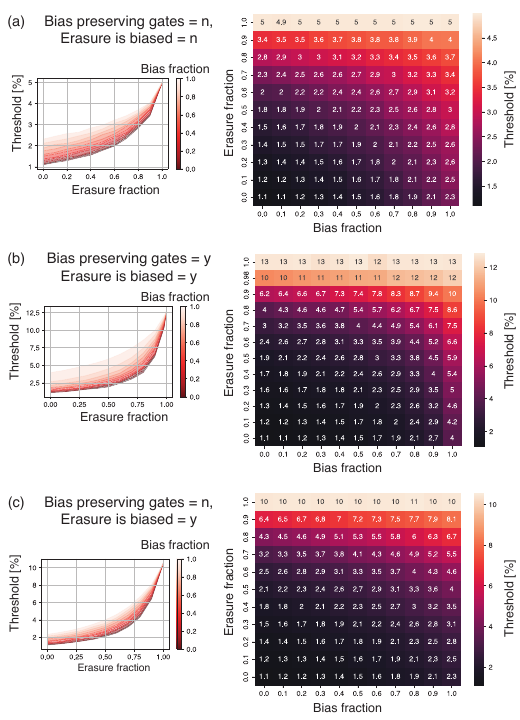}
  \caption{Thresholds for various values of erasure and bias fractions, in cases: (a) without bias preserving gates and erasure is not biased, (b) with bias preserving gate and erasure is biased, and (c) without biased preserving gates and erasure is biased. The heatmaps on the right provide numerical values for the thresholds illustrated on the left plots.}
  \label{fig_supp:erasure_bias_plots}
\end{figure}

We examine the combined influence of bias and erasure on logical-memory performance. 
Using the direct conversion SE method with period 0.25 (the erasure channel) and the first error model described in Appendix~\ref{supp:error_models_loss}, we perform circuit-level simulations of the XZZX surface code for various bias values under both bias-preserving gates and biased-erasure conditions. The data are decoded using our delayed-erasure decoder together with a minimum-weight perfect-matching (MWPM) decoder. The complete results are shown in Fig.~\ref{fig_supp:erasure_bias_plots}. 
We analyze three scenarios: (a) without bias-preserving gates and unbiased erasure, (b) with bias-preserving gates and biased erasure, and (c) with bias-preserving gates and unbiased erasure. 
Numerical threshold values are listed in the tables next to each plot. 
As shown, increasing the erasure fraction influences the thresholds far more strongly than increasing the bias fraction, even in the presence of bias-preserving gates. 
Furthermore, when bias-preserving gates are absent, the effect of bias is reduced, yielding approximately a two-fold improvement in the threshold in the limit of infinite bias (only Z errors).

\subsection{The interplay between loss and biased errors}

\begin{figure}[h]
  \centering
  \includegraphics[width=\columnwidth]{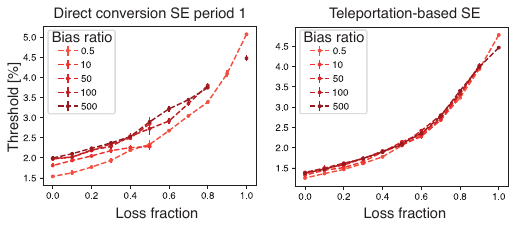}
  \caption{Thresholds for various values of loss fractions and Pauli bias fractions, for methods: (a) Direct conversion SE with period 1, (b) teleportation-based SE. We do not use biased-preserving gates.}
  \label{fig_supp:loss_bias_plots_memory}
\end{figure}

We now turn to simulations that include loss in the presence of biased Pauli errors. 
Using the first error model described in ~\ref{supp:error_models_loss}, we vary both the bias and loss fractions and calculate the threshold for each parameter set.
For the teleportation-based SE method, we employ the XZZX cluster state shown in Fig.~\ref{fig:MBQC_surface_code} (see also Ref.~\cite{claes2023tailored}), without bias-preserving gates. 
For the direct conversion SE method with period 1, we use the surface code, likewise without bias-preserving gates. 
In this regime, we confirm that the regular surface code and the XZZX surface code exhibit equivalent performance under biased noise, as expected.
Circuit-level simulations are performed to estimate the thresholds of each SE method under simultaneous loss and biased Pauli errors. We decode the data using our delayed-erasure decoder combined with an MLE decoder. 
The results are shown in Fig.~~\ref{fig_supp:loss_bias_plots_memory}. 
Consistent with the erasure-and-bias results above, bias increases the threshold by roughly a factor of two in the limit of infinite bias. 
Moreover, increasing the loss fraction has a considerably stronger effect on the threshold than increasing the bias fraction.

\section{Syndrome extraction (SE) methods with loss detection}
\label{app:SE_methods_info}
As explained in the main text, there are multiple approaches to obtaining syndrome measurements and performing QEC, each with its own advantages and disadvantages. Specifically, each method provides different lifecycle lengths and requires varying levels of overhead. Additionally, each method demands different experimental capabilities. Here, we provide further information on each method and the assumptions made in this paper to simulate and compare all methods.

\subsection{Loss detection using physical SWAP SE}
\label{app:SWAP_SE}

This section examines the physical SWAP method for loss detection and correction during rounds of stabilizer measurements. The method integrates stabilizer measurements with loss detection by leveraging the ability to SWAP both the positions and the quantum information of data and measure qubits in each round. Using SSR, it exploits all three measurement outcomes ($\ket{0}$, $\ket{1}$, or loss) to infer both the stabilizer value and the loss status of the data qubits.

\begin{figure}[h]
  \centering
  \includegraphics[width=\columnwidth]{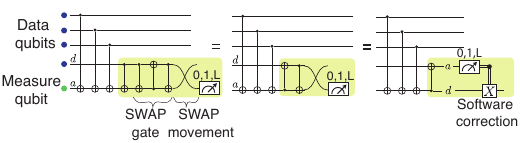}
  \caption{SWAP SE circuit. Left: the original circuit with the four gates for a parity check, plus a SWAP gate and a SWAP movement. Right: an equivalent circuit that uses only four gates and therefore adds no extra gate overhead compared with the conventional SE approach.}
  \label{fig_supp:SWAP_circuit}
\end{figure}

The SWAP method operates by pairing each data qubit with a measure qubit for every stabilizer-check round, as shown in Fig.~\ref{fig_supp:SWAP_circuit}. The measure qubit interacts with its neighboring data qubits through four gates (in the surface-code case). After completing the stabilizer check, the measure and data qubits SWAP their quantum information and physical locations. If the data qubit remains present, the measurement yields $\ket{0}$ or $\ket{1}$, reflecting the stabilizer outcome. If instead the data qubit is missing, the SWAP operation fails, and the measure qubit heralds loss through SSR. This enables immediate detection of data-qubit loss and replacement with a fresh measure qubit.

In a conventional SE round (without the additional SWAP operations), SSR detects loss only on measure qubits, whereas during a SWAP SE round, SSR detects loss only on data qubits. Loss of a measure qubit before the SWAP remains undetected, resulting in replacement of a valid data qubit with a lost one; however, this loss is detected in the subsequent SWAP SE round.

\subsubsection{Optimizing SWAP period}
\label{supp:SWAP_frequency}

\begin{figure}[h]
  \centering
  \includegraphics[width=\columnwidth]{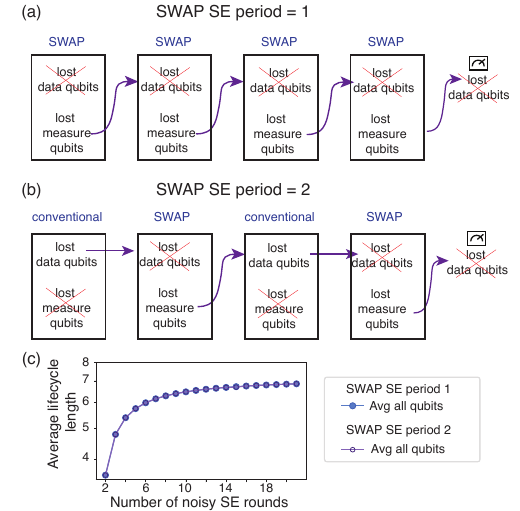}
  \caption{Lifecycle analysis of qubits under SWAP SE with different periods. Panels (a) and (b) depict detected losses (red X marks) via SSR. (a) Continuous SWAP operations (period = 1): each loss of a measure qubit is converted into a data-qubit loss in the next cycle. (b) SWAP SE with period 2: rounds alternate between SWAP and conventional SE. Edge cases (not shown) typically exhibit even longer lifecycles due to the absence of SWAP partners. (c) Average qubit lifecycles for both methods at $d=9$. Both periods exhibit the same average lifecycle length, which approaches 8 in the large-distance limit.}
  \label{supp_fig:SWAP_freq12_lifecycles}
\end{figure}

Each SWAP operation incurs a cost, such as idling errors due to movement. Evaluating different SWAP periods is therefore crucial for identifying the frequency that minimizes SWAP operations in practical settings. Generally, increasing the SWAP period extends the operational lifetime of each qubit. Interestingly, periods of 1 and 2 yield the same average qubit lifecycle, as illustrated in Fig.~\ref{supp_fig:SWAP_freq12_lifecycles}. 
Briefly, with a period of 1, both data and measure qubits have an average lifecycle of $\sim 8$, since they are replaced every round. 
With a period of 2, data qubits have an average lifecycle of $\sim 12$ because they are not subject to loss detection in every round, while during conventional SE rounds the measure qubits have an average lifecycle of only $\sim 4$, as they are not converted into data qubits. This makes a period of 2 competitive with a period of 1 in terms of loss detection while reducing complexity.

\begin{figure}
  \centering
  \includegraphics[width=\columnwidth]{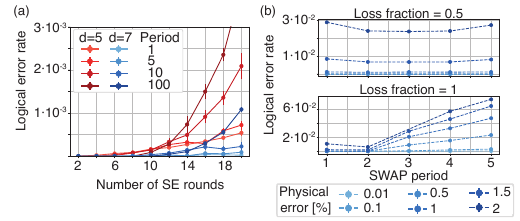}
  \caption{Optimizing the SWAP period in the presence of movement errors. (a) Logical error vs.\ number of SE rounds for SWAP SE with movement errors at a physical error rate of $0.5\%$ and distances $d\in\{5,7\}$. Blue (red) curves show $d=7$ ($d=5$), with a gradient indicating different SWAP periods. A period of 100 (effectively no SWAP loss detection, meaning conventional SE) yields the worst errors. While small periods are competitive, period 1 is not always optimal. (b) Logical error vs.\ SWAP period for different physical error rates, evaluated after 20 SE rounds, at loss fractions $L\in\{0.5,1\}$. The optimal period depends on both the loss fraction and the physical error rate.}
  \label{app_fig:optimal_SWAP_period}
\end{figure}

We incorporate movement-error costs into SWAP logical-memory circuit-level simulations to determine the optimal SWAP period across different loss fractions and physical error rates. According to the data in Fig.~\ref{app_fig:optimal_SWAP_period}, the optimal period depends on the specific error model, enabling reduced experimental demands by selecting the most efficient configuration.

\subsubsection{SWAP SE technical challenges}
\label{Supp_sec:SWAP_technical_challenges}
The strength of the physical SWAP SE method lies in its ability to detect data-qubit loss and replace it with a fresh measure qubit. However, several practical considerations arise:

\begin{itemize}
    \item \textbf{SWAP pairs.} SWAP pairs are organized according to the final gate executed in each round, leaving $O(d)$ qubits along the lattice edge temporarily unpaired. To avoid introducing additional qubits and to ensure that every qubit is paired across rounds, we alternate the gate order between even and odd SWAP rounds. Each SWAP SE round type (even or odd), is characterized by a different ordering of the parity check gates, thus different SWAP pairs. 

    \item \textbf{Movement errors.}
    During the execution of SWAP rounds, the physical movement of qubits introduces an opportunity for error. These errors, termed 'movement errors', arise due to idle errors that occur during the qubit's idle time. 
    Let $p_{\text{idle}}$ denote the per-slot Pauli error probability during idling (aggregated over axes), $\tau$ the duration of one idling slot, and $T$ the total movement (or effective idle) time. The resulting movement-induced error is modeled as
    \begin{equation}
        p_{\text{move}} \;=\; 1 - \bigl(1 - p_{\text{idle}}\bigr)^{T/\tau}.
    \end{equation}
    This expression captures the accumulated effect of independent idle errors across $T/\tau$ slots and is applied uniformly to all qubits that incur the same motion-induced idle time.
\end{itemize}

\subsubsection{Comparing SWAP SE and conventional SE}
SWAP-based SE enables loss detection during each round, but it has disadvantages discussed above—most notably the sacrifice of some syndrome information and the fact that loss can be detected on only one qubit type (data qubit or measure qubit) at a time.

\begin{figure}[t]
  \centering
  \includegraphics[width=\columnwidth]{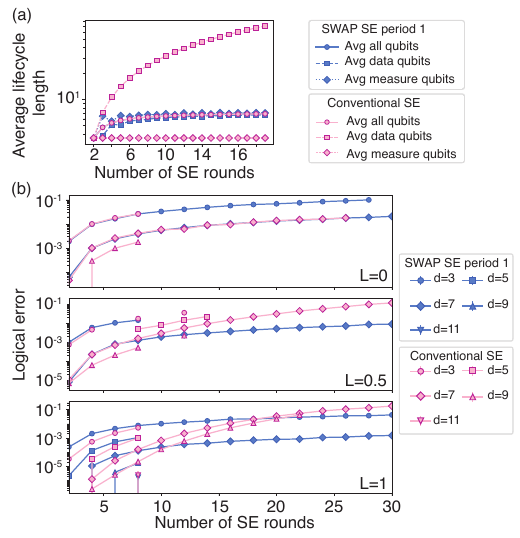}
  \caption{(a) Lifecycle analysis for SWAP SE and conventional SE at distance $d=9$. The average lifecycle over all qubits is equal for both methods. As the number of noisy SE rounds increases, data-qubit lifecycles in conventional SE grow, while in SWAP SE they remain constant. (b) Logical error vs.\ number of SE rounds for SWAP SE and conventional SE at physical error rate $p=1\%$ and various distances. Top to bottom: loss fractions $L=0,\,0.5,\,1$. For $L=1$ (loss-only), SWAP SE becomes beneficial over the conventional SE after approximately six SE rounds.}
  \label{fig_supp:lifecycles_SWAP_None}
\end{figure}

For short circuits (i.e., small total lifecycle lengths), it can be advantageous to avoid SWAP operations altogether. 
Figure~\ref{fig_supp:lifecycles_SWAP_None}(a) compares average lifecycles for data and measure qubits under the two methods. In conventional SE, without the SWAP elements at the end of round, data-qubit lifecycles grow rapidly with the number of noisy SE rounds, while in SWAP-based SE they remain constant. Conventional SE maintains short lifecycles for measure qubits at the expense of longer lifecycles for data qubits; SWAP SE balances both.

We also compare logical performance using circuit-level simulations (Fig.~\ref{fig_supp:lifecycles_SWAP_None}(b)). Both methods use the delayed-erasure decoder developed in this work, and we assume a noiseless first SE round and SSR measurement. As expected, for a small number of rounds, conventional SE outperforms SWAP-based loss detection. The crossover occurs at approximately $k \sim 6$ noisy SE rounds and is roughly distance-independent in the data presented here.

These observations can be understood by linking the numerical trends to the lifecycle behavior: when only a few noisy SE rounds are executed, the longer data-qubit lifecycles induced by SWAP SE do not yet pay off, whereas for deeper circuits SWAP SE’s constant lifecycles provide an advantage. This is consistent with the perspective in Ref.~\cite{fowler2012surface}, where errors on data and measure qubits are effectively categorized into distinct channels and impact thresholds differently.

\subsection{Loss detection using direct conversion SE}
\label{app:FREE_loss_detection_method}

Another method considered in this work is direct loss-to-erasure conversion, applied either after every gate, after multiple gates, or after each SE round. 
In this approach, the experiment detects loss events without requiring additional gates or qubits, instead relying on other hardware capabilities demonstrated in Yb-atom and superconducting-qubit systems~\cite{wu2022erasure, ma2023high, google2024quantum}. This technique requires mid-circuit measurement and replacement of lost qubits, with overheads that depend on the physical implementation.

Since we cannot simulate these experimental overheads, we assume that once a direct conversion occurs, all losses are converted into delayed erasures, corresponding to an experimental platform that successfully detects and replaces each lost qubit.
These simulations therefore provide an upper bound on the threshold and effective distance achievable with mid-circuit erasure-conversion techniques.
We vary the detection and replacement period from $0.25$ (detection and replacement after every gate) to $k \ge 1$ (detection and replacement after every $k$ SE rounds).
We also consider a distinct scenario in which detection is performed more frequently than replacement, since the two processes rely on different experimental capabilities and may require different timescales. 
Specifically, we simulate the case of detection after every gate (period $0.25$) but replacement only at the end of each round (period $1$). We name this approach a direct conversion SE period 1 with loss moment information.

\subsection{Loss detection using teleportation-based SE}
\label{supp:teleportation_based_syndrome_extraction}

In this section, we describe the teleportation-based SE method examined in this work, which employs teleportation of all qubits in each step to both measure stabilizers and detect loss.

\begin{figure}[h]
  \centering
  \includegraphics[width=\columnwidth]{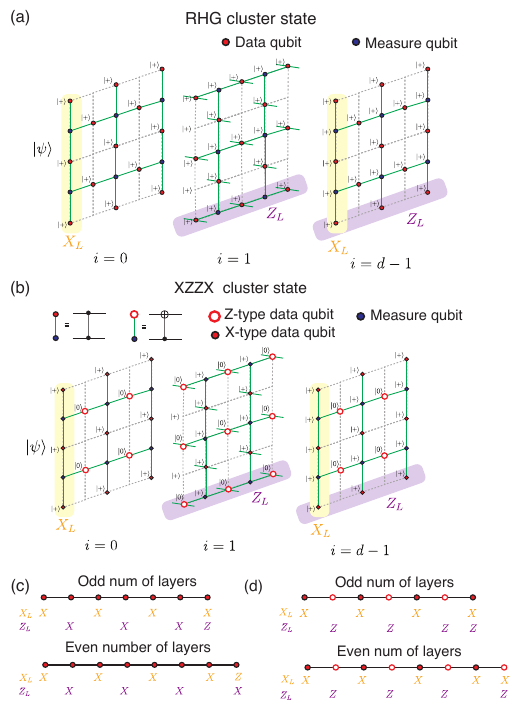}
  \caption{Implementing teleportation-based SE. Construction of the RHG cluster state (a,c) and the XZZX cluster state (b,d) for teleportation-based SE with the surface code.}
  \label{fig:MBQC_surface_code}
\end{figure}

We implement the teleportation-based SE architecture using the foliated surface code, also known as the RHG cluster state~\cite{raussendorf2003measurement, raussendorf2001one, raussendorf2002one}. 
All entangling gates are CZ gates, and qubits are initialized in $\ket{+}$, except for the first layer, which encodes the logical qubit according to the standard surface-code encoding. 
The main-text results correspond to the XZZX cluster state, which, in the absence of bias-preserving gates, reproduces the RHG cluster state.

Figure~\ref{fig:MBQC_surface_code} shows the full construction, including an explicit example of a distance-$d=3$ surface code. 
The figure also illustrates how the logical operators $X_L$ and $Z_L$ propagate through the cluster state, demonstrating both even- and odd-number of layers cases.
Because each layer detects only $X$ or $Z$ errors, we simulate $2d$ layers for all threshold and effective-distance plots presented in this work. 
This approach introduces a space-time overhead corresponding to a $d \times d \times d$ lattice containing $d \cdot (2d^2 + d^2 - 1) = d \cdot (3d^2 - 1)$ physical qubits, with a cluster state generated by 6 time steps.
As in other SE-methods simulations, the first and last layers are considered noiseless.

\subsection{\textit{Modified Steane SE}}
\label{app:Steane_SE}

\begin{figure}[t]
  \centering
  \includegraphics[width=\columnwidth]{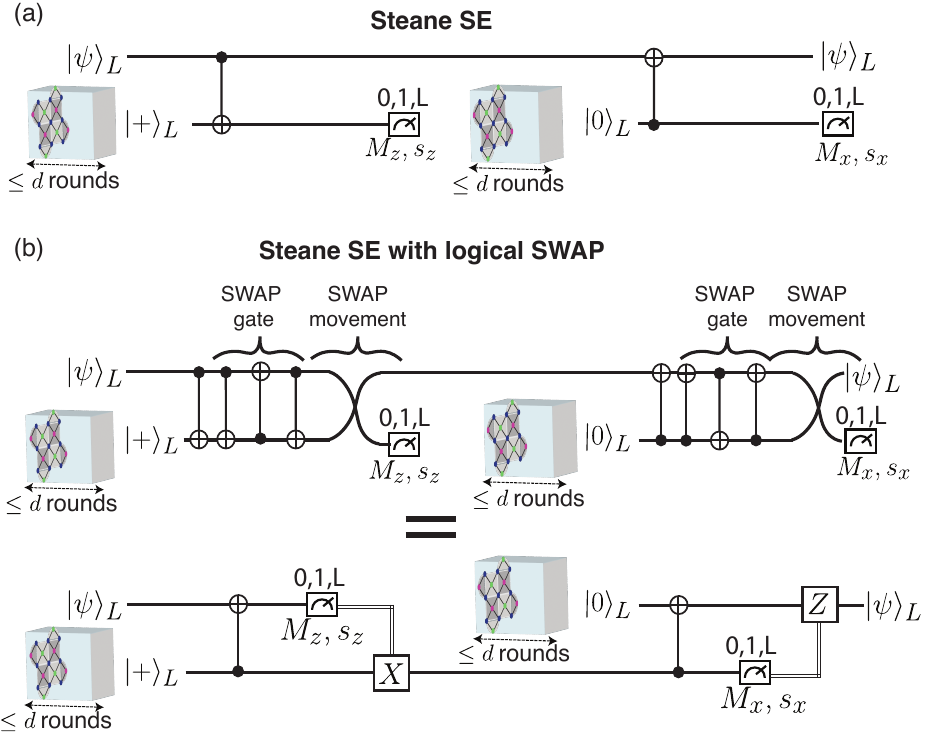}
  \caption{Adjusting the Steane SE method to utilize loss errors. (a) Regular Steane SE circuit and logical-qubit connections. (b) Adjusted Steane SE with a logical SWAP, yielding a Knill-style SE variant that corrects both Pauli and loss errors. Each SSR measurement is used to detect both error types while teleporting quantum information to another logical qubit.}
  \label{fig:Steane_Appendix}
\end{figure}

\subsubsection{Loss detection using Modified Steane SE}

To provide a comprehensive perspective on SE techniques, we include a modified version of Steane SE adjusted for loss errors, although we do not numerically analyze it in the main text. We highlight its conceptual advantages as a promising direction for future studies.
Steane SE uses transversal $CX$ gates applied to fault-tolerant logical measure qubits to extract syndromes and detect Pauli errors~\cite{steane1997active}. 
Transversal gates, applying the same operation across corresponding physical qubits in each code block, ensure that errors propagate predictably from logical data qubits to logical measure qubits.
If the measure qubits are prepared fault-tolerantly with $d$ SE rounds, one round of Steane SE suffices to capture the syndromes in a given basis~\cite{steane1997active}. 
However, conventional Steane SE lacks a native mechanism for detecting loss: losses on data qubits are not observed by the logical measure qubits and remain undetected until the end.

We modify conventional Steane SE by incorporating logical SWAP operations inspired by the physical SWAPs used in SWAP SE, as illustrated in Fig.~\ref{fig:Steane_Appendix}, thereby enabling loss detection through SSR at each logical measurement. 
These logical SWAPs exchange the information of the logical data and measure qubits at each transversal $CX$ gate. This effectively teleports the logical data qubits at every syndrome-extraction step, similar in spirit to Ref.~\cite{knill2005quantum}, and ensures that no physical qubit retains a long lifecycle. 
This modified Steane SE scheme combines the benefits of teleportation-based techniques with the simplicity of transversal operations, minimizing qubit lifecycles while integrating loss detection into the SE process.

A key advantage of Steane SE relative to the other methods studied here is the ability to preselect on the quality of the logical measure-qubit blocks.
To gauge the quality of the prepared measure-qubit blocks, one can employ the logical-gap method for preselection, potentially increasing fidelity without significant overhead~\cite{sahay2022decoder,gidney2023yoked,smith2024mitigating}.
This method uses the syndromes collected during preparation to compute the gap between the probability of the proposed error and that probability conditioned on applying the logical operator.
A large (small) gap indicates that the decoder is confident (not confident) in its correction.
One can then set a cutoff on the gap and keep only measure-qubit blocks whose gap exceeds the threshold.

\subsubsection{Connecting Steane SE with logical SWAP to teleportation-based SE}

\begin{figure}[t]
  \centering
  \includegraphics[width=\columnwidth]{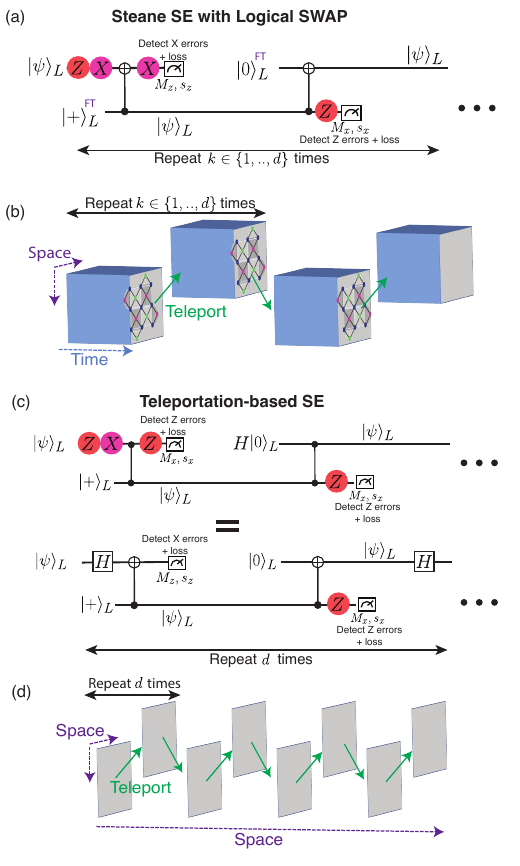}
  \caption{Connection between Steane SE with logical SWAP and teleportation-based SE, revealing an intermediate scheme. (a) Steane SE circuit (without Pauli corrections), showing logical data and measure qubits, error propagation, and detection. This circuit can repeat $k \in \{1,\ldots,d\}$ times; $k=1$ is standard Steane SE, and $k>1$ uses logical measure qubits that are prepared with less rounds and can be preselected. (b) Teleportation illustration of the modified Steane QEC. 
  (c) Teleportation-based SE circuit (without Pauli corrections), equivalent to (a) for $k=d$. 
  (d) Teleportation process in teleportation-based SE. 
  The red and pink circles with $X$ and $Z$ in (a,c) indicate errors and their propagation/detection.}
  \label{supp_fig:connecting_MBQC_and_Steane}
\end{figure}

The modified Steane SE can be viewed as an interpolation between modified conventional SE and teleportation-based SE, similar to the ideas in Ref.~\cite{huang2021between}. 
In the limit of many logical measure qubits, each prepared with a single SE round, the modified Steane SE becomes equivalent to teleportation-based SE (see Fig.~\ref{supp_fig:connecting_MBQC_and_Steane}). 
Conversely, multiple SE rounds can be used to prepare higher-quality logical measure qubits, providing flexibility to match different error models and resource constraints. 
Thus, Steane SE supports any combination between these two options, with $n \in \{1,\ldots,d\}$ preparation rounds per logical measure qubit and $k \in \{1,\ldots,d\}$ rounds of logical teleportation. 
Because modified Steane SE interpolates between modified conventional SE and teleportation-based SE, we expect its performance to qualitatively reflect this interpolation.

\section{Comparing SE Methods Lifecycles and Space-time Overheads}
\label{supp_sec:SE_methods_comparison}


\begin{table*}[t]
    \centering
    \resizebox{\textwidth}{!}{
    \large
    \begin{tabular}{|P{5.5cm}|P{2.8cm}|P{2.8cm}|P{2.8cm}|P{2.8cm}|P{2.8cm}|P{2.8cm}|P{2.8cm}|}
        \hline
        \textbf{Category} 
        & \textbf{Conventional SE} 
        & \textbf{SWAP SE period 1} 
        & \textbf{Direct conversion SE period 2} 
        & \textbf{Teleportation-based SE} 
        & \textbf{Direct conversion SE period 1} 
        & \textbf{Direct conversion SE period 1 + exact loss information} 
        & \textbf{Direct conversion SE period 0.25 + exact loss information} \\
        \hline
        \textbf{Average lifecycle length of data qubits} 
        & $4d$ 
        & 8 
        & 8 
        & 4 
        & 4 
        & 1 (detection) / 4 (replacement) 
        & 1 (detection) / 1 (replacement) \\
        \hline
        \textbf{Average lifecycle length of measure qubits} 
        & 4 
        & 8 
        & 4 
        & 4 
        & 4 
        & 1 (detection) / 4 (replacement) 
        & 1 (detection) / 1 (replacement) \\
        \hline
        \textbf{Space-time overhead} 
        & $\sim 8d^3-4d$ 
        & $\sim 8d^3-4d$ 
        & $\sim 8d^3-4d$ 
        & $\sim 18d^3-6d$ 
        & $\sim 8d^3-4d$ 
        & $\sim 8d^3-4d$ 
        & $\sim 8d^3-4d$ \\
        \hline
        \textbf{Extra experimental requirements} 
        & None 
        & SSR 
        & Erasure conversion~\cite{wu2022erasure, ma2023high}, detection and replacement every 2 SE rounds 
        & SSR 
        & Erasure conversion~\cite{wu2022erasure, ma2023high}, detection and replacement every SE round 
        & Erasure conversion~\cite{wu2022erasure, ma2023high}, detection every gate, and replacement every SE round 
        & Erasure conversion~\cite{wu2022erasure, ma2023high}, detection and replacement every gate \\
        \hline
        \shortstack{\textbf{Threshold for $L=1$ [\%]} \\ (Channel 1)} 
        & NA 
        & $2.27 \pm 0.02$ 
        & $3.98 \pm 0.06$ 
        & $4.78 \pm 0.01$ 
        & $5.07 \pm 0.02$
        & $7.16 \pm 0.01$ 
        & $9.51 \pm 0.04$ \\
        \hline
        \shortstack{\textbf{Effective distance for $d=7$} \\ (Channel 1)} 
        & NA 
        & $6.41 \pm 0.27$ 
        & $6.77 \pm 0.2$ 
        & $7.19 \pm 0.45$ 
        & $6.83 \pm 0.37$ 
        & $6.97 \pm 0.24$ 
        & $7.46 \pm 0.36$  \\
        \hline
        \shortstack{\textbf{Threshold for $L=1$ [\%]} \\ (Channel 2)} 
        & NA 
        & $1.94 \pm 0.01$ 
        & $2.75 \pm 0.07$ 
        & $3.5 \pm 0.05$ 
        & $3.76 \pm 0.01$ 
        & $5.21 \pm 0.01$ 
        & $6.5 \pm 0.01$ \\
        \hline
        \shortstack{\textbf{Effective distance for $d=7$} \\ (Channel 2)} 
        & NA 
        & $6.13 \pm 0.26$ 
        & $5.5 \pm 0.23$ 
        & $6.5 \pm 0.21$ 
        & $6.32 \pm 0.87$ 
        & $6.45 \pm 0.47$ 
        & $7.47 \pm 0.49$ \\
        \hline
    \end{tabular}
    }
    \caption{Comparison of SE methods under various error models. Each column represents a method, and each row describes a property such as average qubit lifecycles, thresholds, effective distance and overheads. The final column estimates performance under very frequent detection and replacement, providing the erasure channel bound. Channel definitions for $L=1$ are as follows. Channel 1: $p' \{L\otimes I, I \otimes L\}$. Channel 2: $\frac{p'}{2} \{L\otimes Z, Z \otimes L, L\otimes I, I \otimes L\}$. $p' = 1 - \sqrt{1-p} \sim p/2 $ where $p$ is the entangling gate physical error rate.}
    \label{table:syndrome_comparison}
\end{table*}

This paper considers multiple approaches for detecting loss and converting to delayed-erasure errors.
As detailed above, we consider methods based on mid-circuit loss detection, known as erasure conversion~\cite{ma2023high, wu2022erasure} in different frequencies (every gate, few gates, rounds), named direct conversion SE.
We also analyze methods that rely solely on SSR, which require adjusting the SE approach to utilize teleportation, such as SWAP SE, and teleportation-based SE.
A further method that can be modified to incorporate teleportation is Steane SE, which we refer to as the modified Steane SE.
Table ~\ref{table:syndrome_comparison} summarizes the average lifecycles, error thresholds, and space-time overheads for each SE method considered numerically in this paper.
The results are presented for multiple error models, including both independent and correlated loss channels between qubits participating in entangling gates.

\section{Numerical simulations methods}

\label{supp:error_models_loss}

\begin{figure}[t]
  \centering
  \includegraphics[width=0.8\columnwidth]{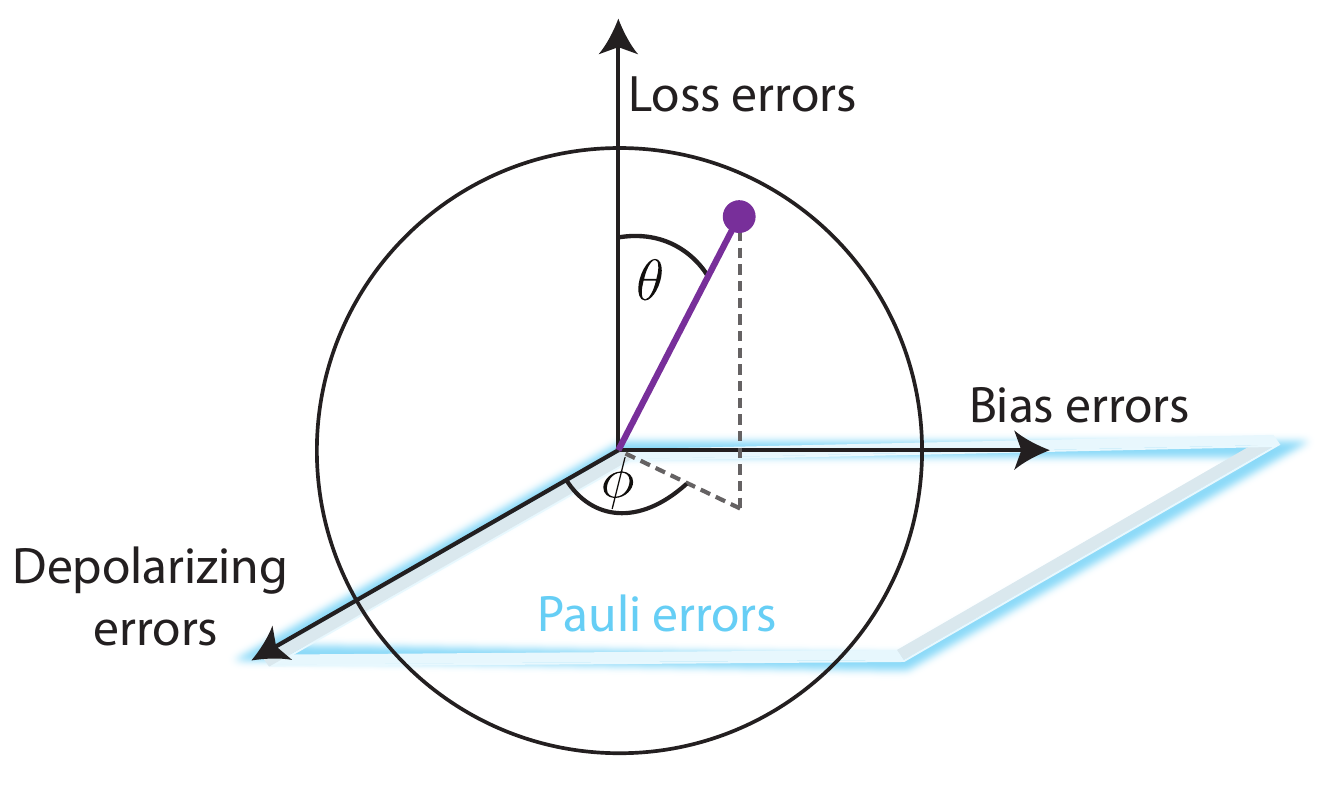}
  \caption{Representation of the error model on a sphere.}
  \label{fig:bloch_sphere}
\end{figure}

We begin by outlining the meta-parameters considered in this work, which are determined by the experimental platform:

\begin{itemize}
    \item \textbf{Bias-preserving gates:} Boolean parameter indicating whether the system’s two-qubit gates preserve error bias (i.e., $Z$ errors remain $Z$). Bosonic systems have demonstrated such gates, whereas atom-based platforms that implement only CZ gates, without direct $CX$ execution, generally do not.

    \item \textbf{Loss is biased:} Boolean parameter specifying whether loss events are biased toward a particular state. For example, in neutral-atom systems where losses predominantly occur from $\ket{1}$, loss is considered biased. Replacing a lost qubit with $\ket{1}$ primarily induces $Z$ errors, reflecting a bias in the effective error channel.

    \item \textbf{Architecture:} Distinguishes between measurement-based quantum computation (used in teleportation-based SE) and circuit-based computation (used in other SE methods).

    \item \textbf{Period of loss-detection rounds:} The number is set by the chosen loss-detection approach.

    \item \textbf{Cost of loss-detection rounds:} Depends on the selected strategy and experimental overhead.

    \item \textbf{QEC code:} Any CSS code can be used; here we focus on the standard and XZZX surface codes.

    \item \textbf{State-selective readout (SSR):} Indicates whether the measurement apparatus can resolve loss events to produce three outcomes: $\ket{0}$, $\ket{1}$, and $\ket{L}$ (loss).
\end{itemize}

Next, we describe the error models that incorporate both loss and biased Pauli errors. 
Throughout this work, errors occur primarily during two-qubit gates, each with probability $p_{\text{CZ}}$ (the physical error parameter reported in the figures). 
We simulate several error models that include correlations between loss and Pauli errors, or between the two qubits involved in a gate.

\textbf{Error model A: Loss-biased Pauli-channel correlations on each qubit.}  
Each qubit experiences an independent error channel with probability $p = 1 - \sqrt{1 - p_{\text{CZ}}}$, ensuring channel normalization. 
Two parameters define the model: bias and loss fraction.  
The relationship can be represented on a unit sphere, as shown in Fig.~\ref{fig:bloch_sphere}. 
The $Z$-axis corresponds to loss errors, which remove the qubit from the computational subspace, while the $XY$-plane corresponds to Pauli errors, which remain within it.  
Pauli errors range from uniform (depolarizing) to single-channel (highly biased) noise, parameterized by angle~$\phi$.  
The angle~$\theta$ links the probabilities of loss and Pauli errors and determines the loss fraction.  
Errors are normalized analogously to a Bloch-sphere vector.  

Given the total error probability~$p$, the likelihood of a loss event is controlled by
\[
L = \cos^2\theta = \frac{p_{\text{loss}}}{p_{\text{loss}} + p_{\text{Pauli}}}.
\]
The biased-Pauli channel is described by
\begin{equation}
    p_x = p_y = \frac{p}{2(1+\eta)}, \qquad
    p_z = \eta(p_x + p_y) = \frac{\eta p}{1+\eta},
\end{equation}
where $\eta$ is the bias parameter related to~$\phi$.  

Normalization check:
\begin{equation}
    P(\text{error}) = 
    \underbrace{L p}_{\text{loss channel}} +
    \underbrace{(1-L)(p_x + p_y + p_z)}_{\text{biased Pauli channel}}
    = p.
\end{equation}
Unless stated otherwise, this is the default model used in our numerical simulations.  
Results based on this model appear in the main text, Table~\ref{table:syndrome_comparison}, and Figs.~\ref{fig_supp:loss_bias_plots_memory}, \ref{app_fig:optimal_SWAP_period}, \ref{fig_supp:lifecycles_SWAP_None}, \ref{fig_supp:LER_num_cycles_all_L}, \ref{fig_supp:thresholds_more_data_L1}, and~\ref{fig_supp:effective_distance}.

\textbf{Error model B: Loss–Pauli correlations between qubits.}  
For $L=1$ (loss-only errors), a loss of one qubit induces a $Z$-error channel on its partner within the two-qubit gate.  
The channel is
\[
\frac{p}{2}\{Z\!\otimes\! L,\; L\!\otimes\! I,\; Z\!\otimes\! L,\; L\!\otimes\! I\},
\]
with $p \approx p_{\text{CZ}}/2$.  
When a qubit is lost, its neighbor experiences a $\tfrac{1}{2}\{Z, I\}$ error channel.  
This is the error model used in Ref.~\cite{yu2024tracking}; our results appear in Table~\ref{table:syndrome_comparison} and Figs.~\ref{fig_supp:thresholds_more_data_L1_loss_Z_corr} and~\ref{fig_supp:effective_distance_loss_Z_corr}.

\textbf{Error model C: Erasure-biased Pauli-channel correlations on each qubit.}  
This model is identical to Error Model A except that loss channels are replaced by perfect erasure channels—representing ideal loss detection and qubit replacement after every gate (direct conversion SE with period 0.25).  
This corresponds to the model used in Ref.~\cite{sahay2023high}; results appear in the main text for direct conversion SE with period 0.25, Table~\ref{table:syndrome_comparison}, Figs.~\ref{fig_supp:erasure_bias_plots},~\ref{fig_supp:thresholds_more_data_L1}, and~\ref{fig_supp:effective_distance}.

\section{Algorithmic procedures}

\begin{figure}[h]
  \centering
  \includegraphics[width=\columnwidth]{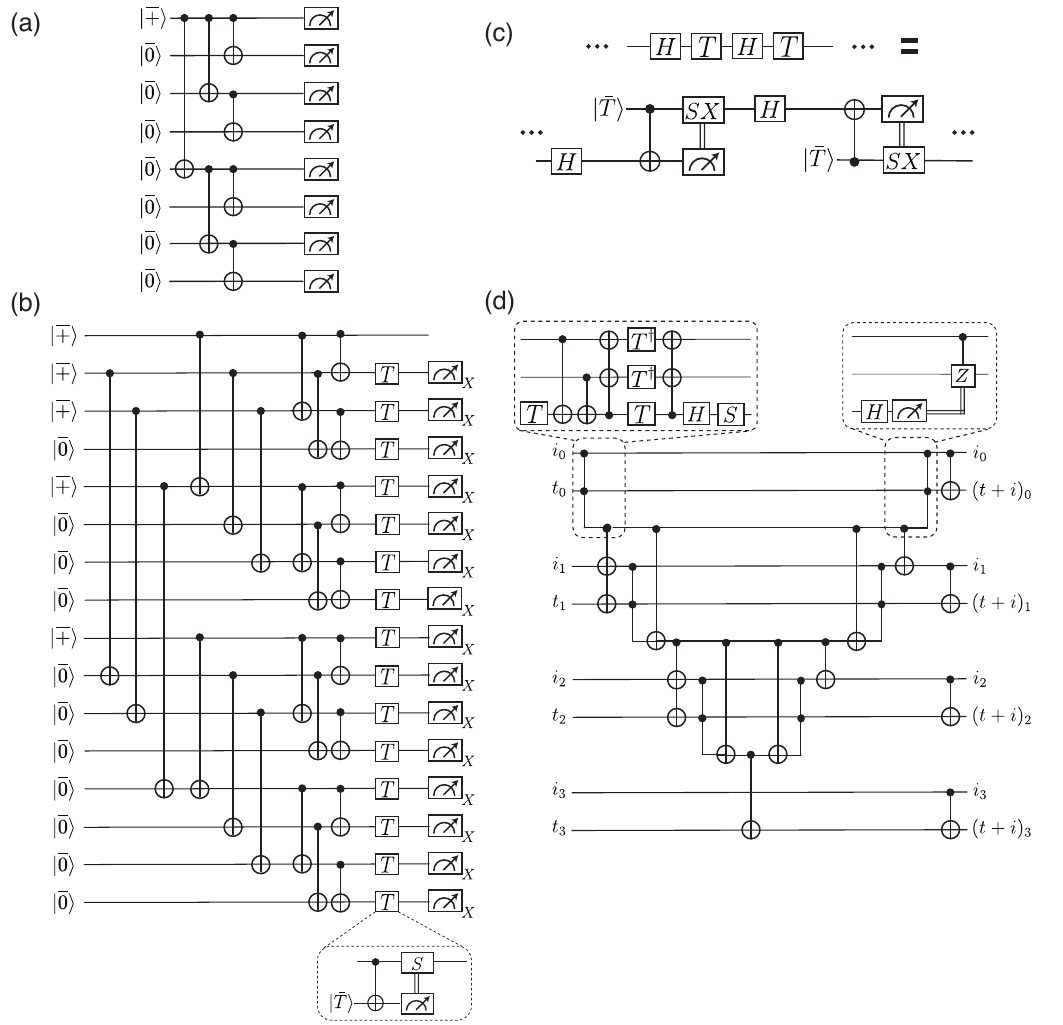}
  \caption{Common logical algorithmic procedures. (a) GHZ state preparation. (b) Magic state distillation circuit. (c) HT decomposition algorithm. (d) Adder circuit.}
  \label{fig:algorithms_lifecycles}
\end{figure}

Here we describe the approach used to calculate the average and maximum lifecycle length for each algorithmic procedure shown in Fig.~\ref{fig:algorithms_lifecycles}. 
For each algorithm, we count the number of noisy SE rounds per logical qubit, from initialization to the final SSR measurement. 
Non-Clifford gates are assumed to be implemented through teleported-gate circuits, which increase the lifecycle but also allow natural loss detection and termination of the lifecycle for one of the logical-data qubits in the teleported-gate circuit.

Inspired by the results in \cite{cain2024correlated, zhou2024algorithmic} and our numerical results presented in the main text, we consider a single SE round for state preparation and a single SE round after every gate.
Accordingly, we count the number of logical $CX$ gates per logical qubit as follows:

\begin{enumerate}
    \item \textbf{GHZ:} We use the logarithmic-depth implementation shown in Fig.~\ref{fig:algorithms_lifecycles}(a). 
    For $n$ qubits, the average number of $CX$ gates per qubit is approximately $\tfrac{\log n}{2}$, and the maximum is $\lfloor \log n \rfloor$.

    \item \textbf{Magic-state distillation:} Shown in Fig.~\ref{fig:algorithms_lifecycles}(b). For each input qubit, there are three or four entangling gates before the logical $T$ gate, which adds another entangling gate and ends the lifecycle of that input qubit. 
    The $T$ gate is performed via teleportation, which measures out the input qubits, giving a total of four or five entangling gates from initialization to measurement. 
    The output qubit passes through three $CX$ gates before being teleported and measured in the next distillation layer. 
    Thus, the average number of $CX$ gates per qubit is~4, with a maximum of~5, independent of the number of layers.

    \item \textbf{HT decomposition algorithm (small-angle synthesis)~\cite{kitaev1997quantum, kitaev2002classical}:} 
    As shown in Fig.~\ref{fig:algorithms_lifecycles}(c), each qubit is measured shortly after initialization, giving a short lifecycle. 
    An input magic state undergoes only two transversal entangling gates before measurement. 
    Assuming it originates from a magic-state distillation circuit, the average lifecycle length is~7, and the maximum is~8.

    \item \textbf{Adder circuit~\cite{gidney2018adder}:} 
    Shown in Fig.~\ref{fig:algorithms_lifecycles}(d).
    The counting differs across qubit rows. 
    Counting entangling gates based on $T$-gate teleportation without SWAP yields an average of~9 in the limit of many logical qubits, with a maximum of~13. 
    Specifically, for the top three qubits, the lifecycle lengths are 6, 6, and 11; for the bottom two, 1 and 2; and for intermediate qubits (in the large-system limit), 8, 7, and 13.
\end{enumerate}

\section{Additional numerical results}

Here we present additional circuit-level simulation results that complement those discussed in the main text. 
All results here are decoded using the delayed-erasure decoder developed in this work, combined with an MLE decoder. 

\subsection{Logical error rate vs.\ loss fraction for various SE methods}

\begin{figure}[h]
  \centering
  \includegraphics[width=\columnwidth]{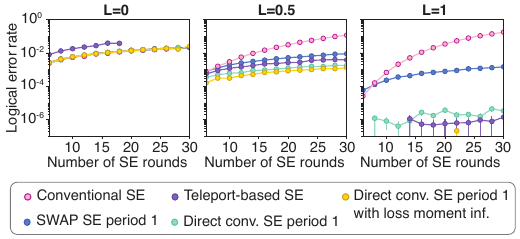}
  \caption{Comparison of different SE methods for various loss fractions~$L$. 
  Logical-memory circuit-level simulations showing the logical error rate as a function of the number of SE rounds. 
  (a)~$L=0$, (b)~$L=0.5$, (c)~$L=1$. 
  The physical error rate is $1\%$ and the code distance is $d=7$.}
  \label{fig_supp:LER_num_cycles_all_L}
\end{figure}

Figure~\ref{fig_supp:LER_num_cycles_all_L} shows the logical error rate as a function of the number of SE rounds for several SE methods. 
Each subplot corresponds to a different loss fraction $L$: 0, 0.5, and~1, respectively. 
For $L=0$, all methods yield comparable results, with teleportation-based SE exhibiting a slightly higher logical error due to additional gate overhead. 
For $L=0.5$, all methods that incorporate loss detection - SWAP SE, teleportation-based SE, and direct conversion SE - show similar performance and support deeper circuits. 
For $L=1$ (loss-only errors), SWAP SE yields higher logical error rates than the other methods because of its longer qubit lifecycles, as explained above.

\subsection{Thresholds as a function of lifecycle length}
\label{supp:decay_vs_lifecycle}
In Fig.~\ref{fig:Phase_diagram}(b) of the main text, we present the thresholds of multiple SE methods as a function of lifecycle length. 
Fitting these data points (excluding SWAP SE) to both exponential and polynomial models shows that the decay follows a polynomial scaling. 
Numerically, we observe a decay proportional to $7/(\text{lifecycle length})^{1/3}$.

\subsection{Thresholds and effective distances for various error models and SE methods}

\begin{figure}[h]
  \centering
  \includegraphics[width=\columnwidth]{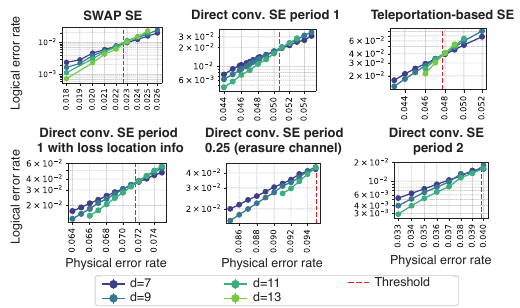}
  \caption{Threshold data: logical error rate vs.\ physical error rate for different SE methods under loss-only errors ($L=1$).
  SE methods considered here:
  SWAP SE, direct conversion SE (period 1), teleportation-based SE, 
  direct conversion SE (period 1) with detection period 0.25 (perfect loss moment information), 
  direct conversion SE (period 0.25) with detection and replacement after every gate (erasure channel), 
  direct conversion SE (period 2). 
  These data are used to determine the threshold for each SE method.}
  \label{fig_supp:thresholds_more_data_L1}
\end{figure}

\begin{figure}[h]
  \centering
  \includegraphics[width=\columnwidth]{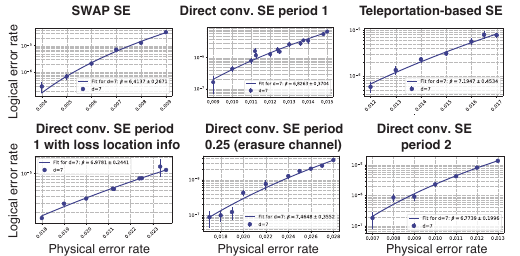}
  \caption{Effective distance data for loss-only errors ($L=1$). 
  SE methods considered here:
  SWAP SE, direct conversion SE (period 1), teleportation-based SE, 
  direct conversion SE (period 1) with detection period 0.25 (perfect loss moment information), 
  direct conversion SE (period 0.25) with detection and replacement after every gate (erasure channel), 
  direct conversion SE (period 2). 
  These data are used to determine the threshold for each SE method.}
  \label{fig_supp:effective_distance}
\end{figure}

\begin{figure}[h]
  \centering
  \includegraphics[width=\columnwidth]{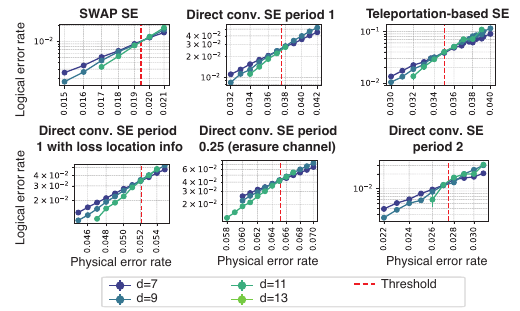}
  \caption{Same as Fig.~\ref{fig_supp:thresholds_more_data_L1}, but for a correlated $Z$–loss error channel (Error Model B), as described in Appendix.~\ref{supp:error_models_loss}.}
  \label{fig_supp:thresholds_more_data_L1_loss_Z_corr}
\end{figure}

\begin{figure}[h]
  \centering
  \includegraphics[width=\columnwidth]{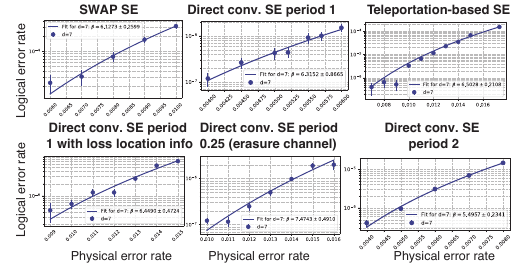}
  \caption{Same as Fig.~\ref{fig_supp:effective_distance}, but for a correlated $Z$–loss error channel (Error Model B), as described in Appendix.~\ref{supp:error_models_loss}.}
  \label{fig_supp:effective_distance_loss_Z_corr}
\end{figure}

Figures~\ref{fig_supp:thresholds_more_data_L1} -~\ref{fig_supp:effective_distance_loss_Z_corr} show the logical error rate as a function of the physical error rate for various SE methods at a loss fraction of~1. 
We consider two different error models: the loss-only model (Error Model A in Appendix~\ref{supp:error_models_loss}, Figs.~\ref{fig_supp:thresholds_more_data_L1},~\ref{fig_supp:effective_distance}) and the correlated loss–$Z$ model (Error Model B in Appendix~\ref{supp:error_models_loss}, Figs.~\ref{fig_supp:thresholds_more_data_L1_loss_Z_corr},~\ref{fig_supp:effective_distance_loss_Z_corr}). 
These results are used to extract both the threshold and effective distance for each error model and SE method.

\subsubsection{Comparison with published errors models}
\label{supp:comparison_with_prev_works}

As a validation step, we benchmark our simulations against other error models, presented in Refs.~\cite{wu2022erasure, sahay2023high, yu2024tracking}. 
First, we perform circuit-level simulations for the case of erasure errors (detected immediately after leaving the qubit subspace), corresponding to direct conversion SE with period 0.25 and perfect operations. 
We then compare our results with those reported in Refs.~\cite{wu2022erasure, sahay2023high}. 
Using the XZZX surface code at various erasure fractions and the two-qubit error model of Ref.~\cite{sahay2023high}, and decoding with our delayed-erasure + MWPM decoder, we reproduce the thresholds reported in Table 1 of Ref.~\cite{sahay2023high}, consistent with the data shown in Fig.~\ref{fig_supp:erasure_bias_plots}.

We further benchmark the teleportation-based SE using the loss–$Z$ correlated error model used in Ref.~\cite{yu2024tracking}. 
Employing our delayed-erasure decoder, we obtain comparable thresholds of approximately 3.5\% for a loss fraction of 1. 
These results are included in Table \ref{table:syndrome_comparison} and Fig.~\ref{fig_supp:thresholds_more_data_L1_loss_Z_corr}.

\section{Random deep logical transversal Clifford circuits}
\label{app:logical_algorithms}

We study deep logical Clifford circuits composed of multiple layers of transversal gates, similarly to Ref.~\cite{cain2024correlated}. 
Each layer consists of single-qubit logical gates from the set $\{X_L, Y_L, Z_L\}$ and transversal CNOT gates applied between random pairs of qubits in a random order. 
After each layer, we perform $n_r$ rounds of SE. 
For $n_r < 1$, a single QEC round is performed after every $1/n_r$ layers, while for $n_r \ge 1$, $n_r$ SE rounds are executed after each layer. 
We use the circuit-level biased-noise Error Model A described in Sec.~\ref{supp:error_models_loss}.
Unless otherwise noted, these noise channels are applied throughout the entire circuit, except during state preparation, the final stabilizer measurements of the last transversal layer, and the final logical stabilizer measurements.

\begin{figure}[h]
  \centering
  \includegraphics[width=\columnwidth]{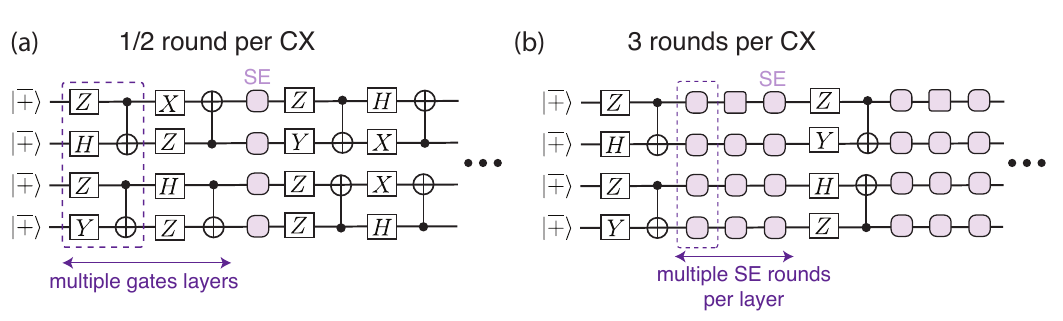}
  \caption{Illustration of deep random Clifford circuits with multiple layers of transversal $CX$ and single-qubit logical gates, including periodic SE rounds. 
  (a)~One SE round every two $CX$ layers ($n_r = 0.5$). 
  (b)~Three SE rounds per $CX$ layer ($n_r = 3$).}
  \label{fig_supp:random_circuit_nr_illustration}
\end{figure}

The main-text figures show the algorithmic logical error probability $P_L$. 
The maximum error $P_{L,\text{max}} = 1 - \tfrac{1}{2^N}$ corresponds to a maximally mixed logical state with $N$ logical qubits.

\nocite{*}
\bibliography{main}


\clearpage
\onecolumngrid

\newpage

\end{document}